\documentclass{article}
\usepackage[T1]{fontenc}
\usepackage[utf8]{inputenc}

\usepackage[english]{babel}
\usepackage{csquotes}
\usepackage{authblk}

\usepackage{lscape}
\usepackage{amsmath}
\usepackage{mathtools}
\usepackage{amsfonts}
\usepackage{amssymb}
\usepackage{algorithm}
\usepackage{algpseudocode}
\usepackage{enumerate}
\usepackage{physics}
\usepackage{hyperref}
\usepackage{xcolor}

\usepackage{natbib}
\usepackage{float}
\usepackage[toc,page]{appendix}
\usepackage[onehalfspacing]{setspace}

\usepackage{adjustbox}
\usepackage{rotating}

\usepackage{booktabs,dcolumn,caption}

\usepackage{subcaption}

\providecommand{\keywords}[1]
{
  \small	
  \textit{Keywords:} #1
}

\newcommand\independent{\protect\mathpalette{\protect\independenT}{\perp}}
\def\independenT#1#2{\mathrel{\rlap{$#1#2$}\mkern2mu{#1#2}}}

\makeatletter
\newcommand*\rel@kern[1]{\kern#1\dimexpr\macc@kerna}
\newcommand*\widebar[1]{%
  \begingroup
  \def\mathaccent##1##2{%
    \rel@kern{0.8}%
    \overline{\rel@kern{-0.8}\macc@nucleus\rel@kern{0.2}}%
    \rel@kern{-0.2}%
  }%
  \macc@depth\@ne
  \let\math@bgroup\@empty \let\math@egroup\macc@set@skewchar
  \mathsurround\z@ \frozen@everymath{\mathgroup\macc@group\relax}%
  \macc@set@skewchar\relax
  \let\mathaccentV\macc@nested@a
  \macc@nested@a\relax111{#1}%
  \endgroup
}
\makeatother

\newcolumntype{L}[1]{>{\raggedright\let\newline\\\arraybackslash\hspace{0pt}}p{#1}}
\newcolumntype{C}[1]{>{\centering\let\newline\\\arraybackslash\hspace{0pt}}p{#1}}
\newcolumntype{R}[1]{>{\raggedleft\let\newline\\\arraybackslash\hspace{0pt}}p{#1}}

\title{Propensity Score Methods for Local Test Score Equating: Stratification and Inverse Probability Weighting}

\author[1]{Gabriel Wallin}
\author[2]{Marie Wiberg}

\affil[1]{School of Mathematical Sciences, Lancaster University}
\affil[2]{Department of Statistics, USBE, Umeå University, Sweden}

\date{}

\begin{document}

\maketitle

\begin{abstract}
In test equating, ensuring score comparability across different test forms is crucial but particularly challenging when test groups are non-equivalent and no anchor test is available. Local test equating aims to satisfy Lord's equity requirement by conditioning equating transformations on individual-level information, typically using anchor test scores as proxies for latent ability. However, anchor tests are not always available in practice. This paper introduces two novel propensity score-based methods for local equating: stratification and inverse probability weighting (IPW). These methods use covariates to account for group differences, with propensity scores serving as proxies for latent ability differences between test groups. The stratification method partitions examinees into comparable groups based on similar propensity scores, while IPW assigns weights inversely proportional to the probability of group membership. We evaluate these methods through empirical analysis and simulation studies. Results indicate both methods can effectively adjust for group differences, with their relative performance depending on the strength of covariate-ability correlations. The study extends local equating methodology to cases where only covariate information is available, providing testing programs with new tools for ensuring fair score comparability.
\end{abstract}


\section{Introduction}

Ensuring fairness in educational testing is a critical issue, especially in the context of large-scale assessments where decisions about examinees are often based on their test scores. The concept of fairness in testing typically revolves around comparability: scores from different forms of the same test should be interchangeable. However, in practice, examinees might take different forms of the test, and ensuring that scores from these forms are comparable is the purpose of test score equating \citep{kolen2014test}. This becomes particularly challenging in the presence of non-equivalent groups, where examinees are not randomly assigned to different forms, leading to systematic differences in the ability distributions of the groups. Such differences must be accounted for in the equating process to ensure that the resulting scores can be meaningfully compared.

In observed-score equating, the goal is to determine comparable scores between two test forms. Depending on the data collection design, different equating methods are used. Two commonly applied designs are the \textit{equivalent groups (EG)} design and the \textit{nonequivalent groups with anchor test (NEAT)} design \citep{kolen2014test, von2004kernel}. In the EG design, two random samples of examinees from the same population take different forms of the test. In contrast, the NEAT design equates the test forms by administering a set of common items, known as the anchor test, to both groups of examinees. The anchor test serves as the link between the two test forms, balancing potential differences between the groups. For further details on these designs, see \citet{kolen2014test}, \citet{von2004kernel}, and \citet{gonzalez2017applying}. 

While the NEAT design is often preferred for equating as it can handle heterogeneous test groups, there are situations where no anchor test is available, particularly in large-scale assessments. For example, the Italian INVALSI test and earlier versions of the Swedish Scholastic Aptitude Test (SweSAT) lacked an anchor test despite having non-equivalent groups \citep{lyren2011consequences}. In such cases, one possible solution is to use a \textit{nonequivalent groups with covariates (NEC)} design, where covariates are used to adjust for group differences. Covariates-based equating methods have been explored in both linear and non-linear equating frameworks \citep{branberg2011observed, wiberg;branberg;2015}. These methods make use of examinees’ background information to balance the groups when anchor items are not available. However, no studies have yet explored the use of covariates for \textit{local equating}, which is the primary focus of this paper.

Common for all test equating methods is that they aim to fulfill certain equating criteria, such as equity and population invariance. However, many of the methods often struggle in real-world scenarios, and part of the reason for this is that they have not taken full consideration of the equating criteria when defining score equivalence \citep{van2010local}. Lord’s equity requirement is particularly important here, as it states that the equated scores from two test forms should be indistinguishable for examinees with the same latent ability \(\theta\) \citep{lord1980applications}. Local equating methods \citep{van2011local, van2010local, wiberg2014local} attempt to satisfy this requirement by conditioning the equating process on examinee-level information, such as an anchor test score or other proxies for latent ability. Both linear and non-linear local equating methods have been proposed \citep{wiberg;vanderlinden;2011}, but those have not yet included the possibility of using covariates.

Using covariates in equating is not a new idea \citep{kolen1990does}, and several researchers have explored their use in matching or as complementary information \citep{cook1990equating, wright1993using, longford2015equating, dorans2008anchor, hsu2002exploring, liou2001estimating}. The propensity score \citep{rosenbaum1983central}, which is a scalar function of the covariates, has also been applied to test equating in a limited capacity. For example, \citet{livingston1990combination} were the first to use propensity scores in test equating, while \citet{paek2006propensity} used propensity score matching to develop a linking relationship between the PSAT and SAT tests. Further, \citet{sungworn2009investigation} used propensity scores based on collateral information to improve poststratification equating. Other researchers, including \citet{moses2010use} and \citet{powers2010impact}, examined the potential for propensity scores to reduce equating biases by combining anchor test scores or demographic information. Recently, \citet{wallin;wiberg;2019} proposed to use propensity scores in the NEC design, however they did not examine the possibility of using it in local equating.

Despite these developments, the use of propensity scores in local equating has not been explored. This paper aims to fill this gap by introducing two new methods: propensity score stratification and inverse probability weighting (IPW) for local equating. Both of these methods have been widely applied in various areas such epidemiology \citep{hernan2006estimating, austin2008performance}, sociology \citep{thoemmes2011systematic, pais2011socioeconomic}, and economics \citep{huber2015causal, vikstrom2017dynamic} to estimate causal effects from observational data. Their popularity stems from their intuitive appeal and their ability to reduce multidimensional covariates to a single scalar summary. Here, these methods are designed to address non-equivalent groups when no anchor test is available, using propensity scores as a proxy for latent ability differences between the test groups to be equated. Propensity score stratification divides examinees into strata based on similar propensity scores, aiming to make the groups balanced in terms of the covariates within each stratum. IPW assigns weights to examinees inversely proportional to their probability of group membership, adjusting for population differences across all score levels. Both methods rely on propensity scores, which represent the conditional probability of group membership based on observed covariates. By using propensity scores as proxies for the unobserved latent ability, we extend local equating to cases where traditional anchor-based methods cannot be applied.

The first method, propensity score stratification, partitions examinees into strata based on their estimated propensity scores. Within each stratum, examinees are assumed to be comparable in terms of their latent ability, allowing for local equating to take place within these balanced groups. This approach ensures that the equating process accounts for the observed differences between groups, conditional on the covariates used to estimate the propensity scores. Stratification on the propensity score is particularly useful in situations where covariates can capture a significant portion of the variability in the latent ability distributions between groups. By equating within each stratum, this method aims to fulfill Lord's equity requirement, ensuring that examinees with similar propensity scores receive equitably equated test scores. The second method, IPW, takes a different approach by assigning weights to each examinee based on the inverse of their propensity score. In this method, the propensity score serves as a weight that adjusts for the differential likelihood of group membership across test forms. By reweighting the sample, IPW effectively creates a pseudo-population in which the distributions of covariates are balanced between the test forms. To evaluate the proposed methods, we first present an empirical study that applies the proposed methods to real test data, illustrating their practical utility. We also conduct an extensive simulation study, where we generated test data under various conditions to evaluate the proposed methods' performance.

The structure of this paper is as follows: Section 2 introduces local equating and the equating estimators considered. Section 3 reports findings from an empirical illustration and section 4 presents the results of the simulation study. Finally, the paper concludes with a discussion of the results and future research directions.

\section{Local Equating}

\subsection{Notation and Background}

Let $X$  and $Y$ denote observed scores on test forms $\mathcal{X}$ and $\mathcal{Y}$, respectively, and let $\theta$ represent the unobserved latent ability influencing responses on both tests. Test score equating aims to find a transformation $\varphi(y)$ that makes scores on test form $\mathcal{Y}$ comparable to scores on test form $\mathcal{X}$. Ideally, this transformation should depend on latent ability $\theta$, resulting in the true ability-specific equating transformation $\varphi_{Y\mid \theta}^*(y)$. The fundamental requirement for such a transformation is given by Lord's equity criterion, which states that the conditional distribution of the equated scores on test form $\mathcal{Y}$ given ability $\theta$ should be identical to the conditional distribution of scores on test form $\mathcal{X}$ given $\theta$. For any ability level $\theta$, this criterion is represented as:
\[
F_{\varphi(Y)\mid \theta}(x) = F_{X\mid \theta}(x) \quad \text{for all } \theta,
\]
where $F_{\varphi(Y) \mid \theta}$ and $F_{X \mid \theta}$ represent the cumulative distribution functions (CDFs) of the equated scores and observed scores, respectively, conditional on $\theta$.

For this criterion to be meaningful, the equating transformation must satisfy certain identification conditions. Specifically, we require:
\[
P(Y \leq y \mid \theta) = P(\varphi(Y) \leq \varphi(y) \mid \theta) \quad \text{for all } y \text{ and } \theta,
\]
This condition ensures that the equating transformation preserves the ordering of scores within each ability level, making it a proper monotonic transformation. Together, Lord's equity criterion and this identification condition imply that the equated scores will be indistinguishable from those on the reference form for examinees at each ability level.

Traditional equating methods use a single transformation across the entire population, which averages across ability levels. There will thus always be a compromise made at each ability level for the equating transformation, which inevitably introduces bias. The resulting equated scores will be influenced by the shape of the ability distribution, leading to population-dependent transformations. Local equating addresses these limitations by focusing on the conditional distributions of the test scores given $\theta$ rather than the marginal distributions. This leads to a \textit{family} of equating transformations, each tailored to a specific level of \( \theta \), ensuring that examinees at different ability levels are equated given their ability level. By conditioning on the ability, local equating yields examinee-level transformations, allowing for score comparability that respects differences across ability levels. To define the target transformation, let \( F_{X \mid \theta}(x) \) and \( F_{Y\mid \theta}(y) \) denote the CDFs for scores \( X \) and \( Y \), conditional on \( \theta \). Local equating defines the transformation as follows:
\[
x = \varphi(y; \theta) = F_{X\mid \theta}^{-1}(F_{Y\mid \theta}(y)),
\]
mapping the percentiles of \( Y \) to those of \( X \) within each ability level \( \theta \).

In practice, however, \(\theta\) is unobserved, and equating methods must rely on proxies for \(\theta\) to approximate these conditional distributions. Common proxies include anchor test scores in the NEAT design or estimates derived from item response theory (IRT) models. The family of local transformations can then be written as:
\[
\varphi(y;z) = F_{X \mid Z = z}^{-1}(F_{Y \mid Z = z}(y)),
\]
where \(Z\) denotes the proxy variable, and \(z\) is a specific proxy value. The validity of using such proxies relies on a surrogate variable assumption, which requires that the proxy captures all relevant information about the relationship between test form assignment and potential outcomes. This can be expressed as:
\[
(X,Y) \independent T \mid Z \iff (X,Y) \independent T \mid \theta,
\]
where $T$ denotes the test form assignment. This condition implies that conditioning on the proxy variable $Z$ is equivalent to conditioning on the latent ability $\theta$ for the purpose of removing selection bias in the test form assignment. When this assumption holds, the proxy-based local equating transformation consistently estimates the true ability-specific transformation. However, the quality of this approximation depends critically on how well the chosen proxy satisfies this surrogacy condition.

By conditioning on proxies for latent ability, local equating improves on traditional methods by reducing bias and population dependence. As demonstrated in previous research \citep{van2010local,wiberg;vanderlinden;2011}, traditional methods like chain equating and poststratification equating can exhibit substantial bias due to their reliance on marginal distributions that do not account for examinee differences in ability. In contrast, local equating maintains Lord's equity criterion by ensuring that the equating transformation is tailored to the specific ability level of each examinee, thereby producing more accurate scores.

In this paper, we assume that the true relationship between the test forms is linear, meaning that the equating process only needs to match the means and variances of the two distributions. Specifically, the family of true equating transformations is given by
\begin{equation}
    \varphi_{Y\mid \theta}^*(y) = \frac{\sigma_{X \mid \theta}}{\sigma_{Y\mid \theta}} (y - \mu_{Y\mid \theta}) + \mu_{X \mid \theta},
\end{equation}
where \(\mu_{X\mid \theta}\) and \(\mu_{Y\mid \theta}\) denote the conditional means of \( X \) and \( Y \) given \( \theta \), and \( \sigma_{X \mid \theta} \) and \( \sigma_{Y \mid \theta} \) are the corresponding conditional standard deviations. In Section \ref{generalisation_equipercentile}, we discuss the proposed method's extension to non-linear methods.

\subsection{Anchor-Based Local Equating}

Anchor-based equating is a well-established method in test equating, relying on a set of common items (the anchor test) administered to all examinees regardless of which test form they take. The anchor test score A can be conceptualized as a noisy measure of the latent ability $\theta$, represented as:
\[
A = g(\theta) + \epsilon,
\]
where $g(\cdot)$ is a monotonic function of $\theta$ and $\epsilon$ represents measurement error satisfying $\epsilon \independent \theta$. This decomposition casts light on a fundamental challenge in anchor-based equating: even if $\theta$ perfectly captures all test form selection effects, the anchor test's effectiveness as a proxy depends on the magnitude of $\epsilon$. When the measurement error is substantial relative to the true score variation, the anchor-based equating transformation may not fully adjust for ability differences between test forms.

The key assumption underlying anchor-based equating is that, despite this measurement error, the anchor score contains sufficient information about ability differences to enable valid equating. Formally, we assume that conditional on the anchor score A, the distributions of X and Y are independent:
\[
(X \independent Y) \mid A.
\]
This assumption implies that any dependence between scores on forms X and Y can be explained by the anchor score, even in the presence of measurement error in A. When this assumption holds, we can define the equating transformation as:
\[
\varphi_A(y) = \frac{\sigma_{X|A=a}}{\sigma_{Y|A=a}}(y - \mu_{Y|A=a}) + \mu_{X|A=a},
\]
where $\mu_{X|A=a}$, $\mu_{Y|A=a}$, $\sigma_{X|A=a}$, and $\sigma_{Y|A=a}$ are the conditional means and standard deviations for each unique anchor score $a$. These moments implicitly account for both the true ability differences and the measurement error in the anchor scores.

The conditional means are estimated by:

\[
\hat{\mu}_{X|A=a} = \frac{1}{n_{X|A=a}} \sum_{i: A_i = a, T_i = 0} X_i \quad \text{and} \quad \hat{\mu}_{Y|A=a} = \frac{1}{n_{Y|A=a}} \sum_{i: A_i = a, T_i = 1} Y_i,
\]
where $T_i = 0$ if subject $i$ took test form $\mathcal{X}$, $T_i = 1$ if subject $i$ took form $\mathcal{Y}$, $n_{X|A=a}$ and $n_{Y|A=a}$ are the number of examinees with anchor score $a$ in test forms $\mathcal{X}$ and $\mathcal{Y}$, respectively. The conditional standard deviations for each anchor score $a$ are calculated as:

\[
\hat{\sigma}_{X|A=a} = \sqrt{\frac{1}{n_{X|A=a} - 1} \sum_{i: A_i = a, T_i = 0} (X_i - \hat{\mu}_{X|A=a})^2} \quad \text{and} \quad \hat{\sigma}_{Y|A=a} = \sqrt{\frac{1}{n_{Y|A=a} - 1} \sum_{i: A_i = a, T_i = 1} (Y_i - \hat{\mu}_{Y|A=a})^2}.
\]

This local approach to calculating standard deviations ensures that the equating transformation adapts to the variability specific to each anchor score, rather than pooling data across all anchors.

This method provides a straightforward approach to equating when a common set of items is available. However, its effectiveness relies heavily on the quality of the anchor test and its relationship to the main test forms. In our study, we will use this method as a benchmark for comparing the performance of the propensity score-based methods.

\section{Propensity Score-Based Equating}

When no anchor test is available, but covariates are present, equating may be performed using propensity scores. The propensity score \( \pi \) is defined as the probability that an examinee was assigned to test form $\mathcal{X}$ (or $\mathcal{Y}$, whichever is set to the active treatment), conditional on a set of covariates \( C_1, C_2, \dots, C_k \):
\begin{equation}
    \pi = P(T = 1 \mid C_1, C_2, \dots, C_k).
\end{equation}
In this case, the propensity score serves as a proxy for \( \theta \) differences between the groups, and we will present two approaches for using propensity scores for local equating: stratification and IPW.

\subsection{Propensity Score Stratification}

Propensity score stratification is a useful method used in observational studies to reduce bias \citep{rosenbaum1984reducing}. This method aims to create subgroups of subjects with similar propensity scores, thereby approximating the balance achieved by randomization in experimental studies. The propensity score serves as a balancing score: when subjects are grouped based on similar propensity scores, the distribution of observed baseline covariates is expected to be similar between treated and untreated subjects within each stratum, mimicking a randomized experiment within these subgroups.

Let $\pi(\mathbf{C}_i)$ denote the propensity score for examinee $i$ with covariate vector $\mathbf{C}_i = (C_{i1},\cdots, C_{ik})^\top$, estimated via logistic regression. The sample is partitioned into $K$ strata based on the quantiles of the estimated propensity scores, with stratum $k$ denoted by $S_k$, where $k = 1, \ldots, K$. This quantile-based stratification ensures approximately equal numbers of examinees across strata. 

The effectiveness of stratification relies on achieving balance within each stratum. Within each stratum $S_k$, we require:
\[
P(T=1 \mid \mathbf{C}, S_k) = P(T=1 \mid S_k),
\]
indicating that, conditional on stratum membership, the probability of test form assignment becomes independent of the covariates. When this balance condition holds, examinees within each stratum have similar distributions of covariates regardless of their test form assignment. The choice of the number of strata $K$ involves a bias-variance trade-off. Under suitable regularity conditions, the bias of the stratified estimator decreases with the number of strata at a quadratic rate, suggesting that increasing the number of strata can substantially reduce bias \citep{cochran1968effectiveness, imbens2015causal}. However, this theoretical advantage must be balanced against the need for sufficient sample size within each stratum to ensure stable estimation of the conditional moments. In practice we chose $K$ for which the absolute standardized mean differences for all covariates fell below a commonly used threshold while within-stratum sample sizes remained adequate, indicating that the accompanying variance inflation was modest relative to the gain in bias reduction.

Within each stratum $S_k$, we calculate form-specific means and standard deviations:

\[
    \mu_{k,X} = \frac{1}{n_{k,X}} \sum_{i \in S_k} (1-T_i)X_i, \quad \mu_{k,Y} = \frac{1}{n_{k,Y}} \sum_{i \in S_k} T_i Y_i,
\]

\[
    \sigma_{k,X} = \sqrt{\frac{1}{n_{k,X} - 1} \sum_{i \in S_k} (1-T_i)(X_i - \mu_{k,X})^2},
\]

\[
    \sigma_{k,Y} = \sqrt{\frac{1}{n_{k,Y} - 1} \sum_{i \in S_k} T_i(Y_i - \mu_{k,Y})^2},
\]

\noindent where $n_{k,X}$ and $n_{k,Y}$ denote the number of examinees taking forms $\mathcal{X}$ and $\mathcal{Y}$ in stratum $k$, respectively. Under the balance condition, these within-stratum moments consistently estimate the conditional moments of the test score distributions. The stratum-specific equating transformation is then defined as:

\[
    \varphi_k(y) = \frac{\sigma_{k,X}}{\sigma_{k,Y}} (y - \mu_{k,Y}) + \mu_{k,X}.
\]

This yields a family of local equating functions, where scores are transformed according to the stratum-specific parameters. 

\subsubsection{Assumptions and Properties}

The validity of the stratified propensity score approach relies on several assumptions:

1. \textit{Positivity}: Each examinee must have a non-zero probability of taking either test form, ensuring that $0 < \pi(C) < 1$ for all examinees.

2. \textit{Unconfoundedness}: Conditional on the observed covariates $C$, test form assignment is independent of potential outcomes:
\[
(X, Y) \independent T \mid C.
\]

3. \textit{Correct Model Specification}: The estimated propensity scores must be consistent for the true conditional probability of test form assignment:
\[
\hat{\pi}(C) \xrightarrow{p} P(T = 1 \mid C).
\]

4. \textit{Adequate Overlap}: Within each stratum, there must be sufficient representation of both test forms to enable meaningful comparisons.

Under these assumptions, the consistency of the stratum-specific estimators can be established. As the sample size increases, the estimated means and standard deviations converge in probability to their population counterparts, and by the continuous mapping theorem, the estimated equating transformation within each stratum converges to the true stratum-specific equating relationship:

\[
\hat{\varphi}_k(y) \xrightarrow{p} \varphi_k^*(y),
\]

where $\varphi_k^*(y)$ represents the true equating transformation within stratum $k$.

The stratification approach offers several advantages. First, it allows for heterogeneous equating relationships across different regions of the propensity score distribution. Second, the quantile-based stratification ensures balanced stratum sizes. However, the method's effectiveness depends critically on proper covariate selection, careful diagnostic assessment of balance within strata, and sufficient sample sizes to support local estimation.

\subsection{Inverse Probability Weighting}

IPW is a statistical technique used in causal inference and observational studies to adjust for confounding and selection bias. The method was first introduced by \citet{horvitz;thompson;1952} in the context of survey sampling and has since been widely applied in various fields, including epidemiology, economics, and social sciences. The core idea of IPW is to create a pseudo-population where the distribution of confounding variables is balanced between treatment groups (or, in our case, test forms). This is achieved by assigning weights to individual observations, with the weights being inversely proportional to the probability of receiving the treatment (or taking a particular test form) given the observed covariates.

In a general setting, let $T$ be a binary treatment indicator, $Y(1)$ the potential outcome under treatment, $Y(0)$ the potential outcome under control, $Y$ the observed outcome, and $C$ be a vector of covariates. The average treatment effect (ATE) can be estimated using IPW as follows:

\begin{equation}
    \text{ATE} = E[Y(1) - Y(0)] = E\left[\frac{TY}{\pi(\mathbf{C})} - \frac{(1-T)Y}{1-\pi(\mathbf{C})}\right].
\end{equation}

In the context of test equating, we adapt this general framework to address the challenge of comparing scores from different test forms. 

\subsubsection{IPW for Test Equating}

As a second method to use propensity scores, we propose a stratified inverse‐probability‐weighting approach for local equating. First, we estimate propensity scores \(\pi(\mathbf{C}_i)\) via logistic regression on a set of observed covariates \(\mathbf{C}_i\). We then partition the sample into \(K\) strata according to the quantiles of the estimated \(\pi(\mathbf{C}_i)\). Within each stratum \(k\), we define stabilized inverse‐probability weights as

\[
  w_{ki}
  =
  \begin{cases}
    \dfrac{p_{k,Y}}{\pi(\mathbf{C}_i)}, & T_i = 1,\\[1em]
    \dfrac{1 - p_{k,Y}}{1 - \pi(\mathbf{C}_i)}, & T_i = 0,
  \end{cases}
\]

where
\[
  p_{k,Y} = P\bigl(T_i = 1 \mid S_i = k\bigr)
\]
is the proportion of examinees in stratum \(k\) who took form \(Y\) (recall \(T_i = 1\) denotes form \(Y\) and \(T_i = 0\) denotes form \(X\)), and \(S_i\) denotes the stratum membership for examinee \(i\). To mitigate the impact of extreme weights, we apply symmetric truncation\footnote{Because the propensity-score stratification procedure already limits the influence of extreme \(\hat{\pi}(\mathbf{C}_i)\) by pooling similar scores into homogeneous strata, no additional weight truncation is required there; the two methods are therefore comparable with respect to the treatment of extreme propensities.} at specified quantiles:

\[
    w_{ki}^* = \begin{cases}
        q_{\alpha/2} & \text{if } w_{ki} < q_{\alpha/2} \\
        w_{ki} & \text{if } q_{\alpha/2} \leq w_{ki} \leq q_{1-\alpha/2} \\
        q_{1-\alpha/2} & \text{if } w_{ki} > q_{1-\alpha/2}
    \end{cases},
\]
\noindent where $q_{\alpha/2}$ and $q_{1-\alpha/2}$ are the $\alpha/2$ and $(1-\alpha/2)$ quantiles of the weights within each stratum, respectively, and $\alpha$ is a small positive number (e.g., 0.01).

Within each stratum $k$, we calculate the weighted moments for each test form. The weighted means are given by:

\[
    \hat{\mu}_{k,\mathcal{X}}^w = \frac{\sum_{i \in S_k} w_{ki}^* (1-T_i) X_i}{\sum_{i \in S_k} w_{ki}^* (1-T_i)}, \quad \hat{\mu}_{k,\mathcal{Y}}^w = \frac{\sum_{i \in S_k} w_{ki}^* T_i Y_i}{\sum_{i \in S_k} w_{ki}^* T_i},
\]
and the weighted standard deviations are:

\[
    (\hat{\sigma}_{k,\mathcal{X}}^w)^2 = \frac{\sum_{i \in S_k} w_{ki}^* (1-T_i) (X_i - \hat{\mu}_{k,X}^w)^2}{\sum_{i \in S_k} w_{ki}^* (1-T_i)},
\]
\[
    (\hat{\sigma}_{k,\mathcal{Y}}^w)^2 = \frac{\sum_{i \in S_k} w_{ki}^* T_i (Y_i - \hat{\mu}_{k,\mathcal{Y}}^w)^2}{\sum_{i \in S_k} w_{ki}^* T_i}.
\]

The IPW equating transformation is then defined locally within each stratum as:

\[
    \hat{\varphi}_{k,IPW}(y) = \frac{\hat{\sigma}_{k,\mathcal{X}}^w}{\hat{\sigma}_{k,\mathcal{Y}}^w} (y - \hat{\mu}_{k,\mathcal{Y}}^w) + \hat{\mu}_{k,\mathcal{X}}^w.
\]

This yields a family of stratum-specific equating functions that transform scores on test form $\mathcal{Y}$ to the scale of test form $\mathcal{X}$, accounting for potential differences in the populations taking each test form through both stratification and within-stratum weighting.

\subsubsection{Assumptions and Properties}

As with the propensity score stratification method, the stratified IPW approach to local equating relies the assumptions of positivity, unconfoundedness, and correct model specification. Under these assumptions, the consistency of the stratum-specific estimator can be established. For any stratum $k$, the weighted means converge in probability to their respective population parameters, and similarly, the weighted standard deviations converge to their population counterparts. By the continuous mapping theorem, it follows that within each stratum, the IPW equating transformation converges to the true equating relationship:
\[
\hat{\varphi}_{k,IPW}(y) \xrightarrow{p} \varphi_k^*(y),
\]
where $\varphi_k^*(y)$ is the true equating transformation for stratum $k$.

The IPW-stratified approach offers several advantages over global weighting. First, it allows for heterogeneous equating relationships across different regions of the propensity score distribution, capturing potential effect modification by covariates. Second, by conducting weight truncation within strata, we can better control the influence of extreme weights while preserving local equating relationships. However, this flexibility comes at the cost of requiring sufficient sample sizes within each stratum to ensure stable estimation of the local equating functions.

\section{Model Generalisations}
The proposed framework for equating using propensity scores can be extended in several ways to accommodate various test structures and equating requirements. Here, we present some generalizations that broaden the applicability of our method.

\subsection{Equipercentile Equating}\label{generalisation_equipercentile}

While our primary focus has been on linear equating, the framework can be extended to equipercentile equating. Equipercentile equating assumes that scores on different forms of a test should have the same percentile ranks in the populations of examinees taking each test form. Let \(F_X\) and \(F_Y\) denote the cumulative distribution functions of scores on test forms \(\mathcal{X}\) and \(\mathcal{Y}\), respectively. The equipercentile equating transformation is defined as
\[
  \varphi_{EP}(y) = F_X^{-1}\bigl(F_Y(y)\bigr).
\]
To implement this within our IPW framework, we estimate \(F_X\) and \(F_Y\) using weighted empirical distribution functions. Recall that \(T_i = 1\) for form \(\mathcal{Y}\) and \(T_i = 0\) for form \(\mathcal{X}\). We then define
\[
  \hat{F}_X(x)
    = \frac{\displaystyle\sum_{i=1}^n w_i\,(1 - T_i)\,I\bigl(X_i \le x\bigr)}
           {\displaystyle\sum_{i=1}^n w_i\,(1 - T_i)},
  \quad
  \hat{F}_Y(y)
    = \frac{\displaystyle\sum_{i=1}^n w_i\,T_i\,I\bigl(Y_i \le y\bigr)}
           {\displaystyle\sum_{i=1}^n w_i\,T_i},
\]
where \(I(\cdot)\) is the indicator function and \(w_i\) are the stabilized inverse‐probability weights.
It's worth noting that linear equating can be viewed as a special case of equipercentile equating. When a kernel function \citep{von2004kernel,wiberg;etal;2025} is used to continuize the test score distributions, linear equating emerges as the limiting case of equipercentile equating as the smoothing parameter tends to infinity. Formally, Let \(k(\cdot)\) be a symmetric, integrable kernel density with cumulative distribution function \(K(u)=\int_{-\infty}^{u} k(v)\,dv\).
For a bandwidth \(h>0\) define the scaled (integrated) kernel
\(K_h(u)=K(u/h)\).
Using the stabilized inverse-probability weights \(w_i\) introduced earlier,
we can estimate the smoothed distribution functions by   
\[
  \hat F_{X,h}(x)=
  \frac{\displaystyle\sum_{i=1}^{n} w_i\,(1-T_i)\,K_h\!\bigl(x-X_i\bigr)}
       {\displaystyle\sum_{i=1}^{n} w_i\,(1-T_i)}, 
  \qquad
  \hat F_{Y,h}(y)=
  \frac{\displaystyle\sum_{i=1}^{n} w_i\,T_i\,K_h\!\bigl(y-Y_i\bigr)}
       {\displaystyle\sum_{i=1}^{n} w_i\,T_i}.
\]
 
If we define the smoothed version of the equipercentile equating transformation as:
$$
\varphi_{EP,h}(y) = F_{X,h}^{-1}(F_{Y,h}(y)),
$$
then:
$$
\lim_{h \to \infty} \varphi_{EP,h}(y) = \frac{\sigma_X}{\sigma_Y}(y - \mu_Y) + \mu_X
$$

This limit is taken at the population level (hence deterministic); when the population moments are replaced by their sample estimates, the same result holds in probability.

This relationship highlights the flexibility of our framework in accommodating different equating methods.

\subsection{Mixed-Format Tests}
Our framework can be extended to handle mixed-format tests, including not only binary items but also nominal or ordinal items. Let $\mathbf{X} = (X_1, \ldots, X_p)$ and $\mathbf{Y} = (Y_1, \ldots, Y_p)$ represent the item responses for test forms $\mathcal{X}$ and $\mathcal{Y}$, respectively, where each item can have a different number of response categories.
For nominal items, we can use multinomial logistic regression to model the propensity scores:
$$
\log\left(\frac{P(T_i = 1 \mid \mathbf{C}_i)}{P(T_i = 0 \mid \mathbf{C}_i)}\right) = \beta_0 + \boldsymbol{\beta}^T \mathbf{C}_i
$$
For ordinal items, we can employ ordinal logistic regression:
$$
\log\left(\frac{P(T_i \leq j \mid \mathbf{C}_i)}{P(T_i > j \mid \mathbf{C}_i)}\right) = \alpha_j - \boldsymbol{\beta}^T \mathbf{C}_i, \quad j = 1, \ldots, J-1
$$
where $J$ is the number of ordinal categories, and $T_i$ denotes the categorical treatment assignment.
The equating transformation for mixed format items can be defined separately for each item type and then combined to form an overall equating transformation. For example, for a test with both dichotomous and polytomous items, we might define:
$$
\hat{\varphi}_{mixed}(y) = \hat{\varphi}_{dichot}(y_{dichot}) + \hat{\varphi}_{poly}(y_{poly})
$$
where $\hat{\varphi}_{dichot}$ and $\hat{\varphi}_{poly}$ are equating transformations for the dichotomous and polytomous parts of the test, respectively.

\section{Empirical Analysis}

To empirically illustrate the proposed local equating methods and compare with existing methods, two consecutive forms from the Swedish Scholastic Assessment Test (SweSAT) were used. SweSAT is a paper and pencil test with 160 multiple-choice binary-scored items, and it consists of a quantitative section of 80 items and a verbal section of 80 items that are equated separately. The SweSAT is given twice a year and is a college admission test. Since 2011, an anchor test is included but previously, different groups with specific values on their covariates have been a major part of the equating process. For details about previous used equating methods for the SweSAT, see \citet{lyren2011consequences}. Note, although anchor tests are used in the equating, covariates are still important to examine and use in the equating process to adjust for nonequivalent test taker groups. The job market highly influences which test takers that take the test and how large the test taking group is for any given year. If there is an economic recession and the unemployment rate is high, more test takers (with a diverse background) tend to take the SweSAT as opposed to when the unemployment rate is low. Furthermore, \cite{lyren2011consequences} found clear signs that the equivalent groups design assumptions were violated for the SweSAT. The empirical study was carried out in R \citep{R;2024}.

\begin{figure}[]
    \centering
        \includegraphics[width=0.8\textwidth]{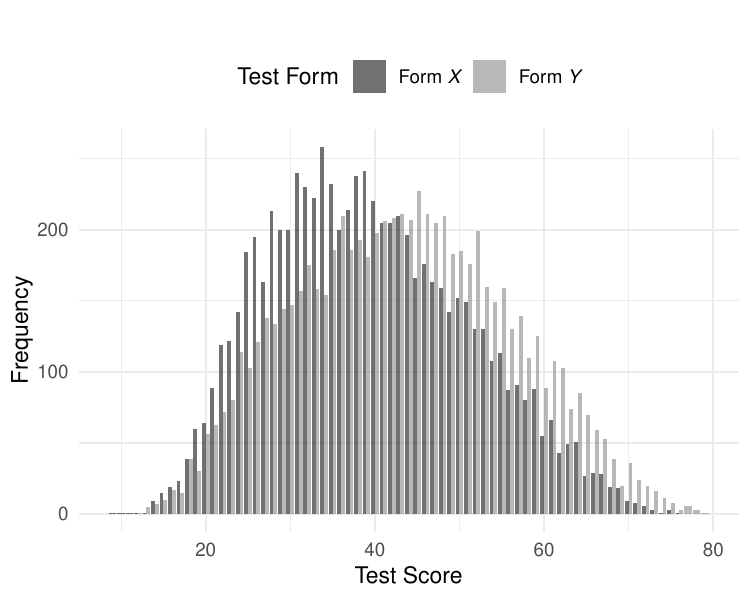}
        \caption{The test score distributions for the analysed SweSAT data.}
        \label{fig:score_distributions}
\end{figure}

The quantitative new test form $\mathcal{X}$ was equated to the old quantitative test form $\mathcal{Y}$. To facilitate comparisons with results using a NEC design, the same data material was used as in Wiberg and Bränberg (2015). The used sample included examinees who had taken both the $\mathcal{X}$ and $\mathcal{Y}$ test forms. From these two test forms a 24-item anchor test was constructed from a selection of 12 items from each administration. Note, if an anchor test is internal or external it is irrelevant to the statistical nature of the equating procedure \citep{van2010local,wiberg;vanderlinden;2011}, thus this issue does not need to be addressed further here.
There were 14,644 examinees who took both test forms. Two equally sized samples of 7,322 examinees for each test were used as the group was divided in half. The score distributions are displayed in Figure \ref{fig:score_distributions}.

In this analysis, we made use of the covariates gender (0 = female, 1 = male), age, and test scores from the verbal section (range: 0–80). A summary of these covariates are given in Table \ref{tab:test_form_stats}. The choice of used covariates depended both on availability of covariates from SweSAT administrations, and the fact that these covariates have been used successfully when calculating propensity scores in previous studies, e.g. \citet{wallin;wiberg;2019}.
\begin{table}[]
\centering
\caption{Summary Statistics and Correlations for Test Forms}
\begin{tabular}{lrrrr}
\hline
& Verb & Age & Gender & Anchor \\
\hline
Test Form $\mathcal{Y}$ correlation & 0.48 & $-$0.14 & 0.26 & 0.81 \\
Test Form $\mathcal{X}$ correlation & 0.52 & $-$0.13 & 0.28 & 0.81 \\
Summary measure$^a$ & 43.91 (39.35) & 1 (1) & 0.42 (0.53) & 12.17 (10.55) \\
Variability measure$^b$ & 12.08 (11.56) & 2 (2) & 0.49 (0.50) & 4.59 (4.64) \\
\hline
\end{tabular}

\smallskip
\begin{minipage}{.95\textwidth}
\small
\textit{Note.} $^a$Values in parentheses correspond to test form $\mathcal{Y}$. 
$^b$For the Age variable, the statistics presented are median and quartile deviation. Age correlations are Spearman coefficients, and Gender correlations are point-biserial. All other measures are means and standard deviations.
\end{minipage}
\label{tab:test_form_stats}
\end{table}
To implement the stratification, we evaluated different numbers of strata through covariate balance diagnostics rather than traditional model fit assessments. The absolute standardized mean difference (ASMD) served as our primary balance metric:
\[
\text{ASMD} = \frac{\left| \bar{C}_X - \bar{C}_Y \right|}{\sqrt{\frac{\sigma_X^2 + \sigma_Y^2}{2}}},
\]
with values below 0.1 indicating satisfactory balance between test forms within a stratum. Through iterative evaluation, we determined that 20 strata provided the best balance while maintaining adequate sample sizes within each stratum. The observed ASMDs across strata ranged from 0.008 to 0.276 for verbal ability, 0.001 to 0.282 for age, and 0.002 to 0.386 for gender. Satisfactory balance (ASMD < 0.1) was achieved in 30\%, 75\%, and 45\% of strata for these covariates, respectively. Complete ASMD values for all covariates and strata are provided in Table \ref{tab:asmd_values}. To ensure the robustness of our results, we conducted sensitivity analyses by varying the number of strata, confirming that slight changes in stratification did not substantially impact the equated scores.

\begin{table}[]
\centering
\caption{Absolute Standardized Mean Differences (ASMD) by Stratum and Covariate.}
\begin{tabular}{cccc}
\toprule
  Stratum & Age & Verb & Gender\\
\midrule
1 & 0.19 & 0.05 & 0.39\\
2 & 0.14 & 0.15 & 0.34\\
3 & 0.13 & 0.12 & 0.31\\
4 & 0.18 & 0.03 & 0.23\\
5 & 0.11 & 0.04 & 0.18\\
\addlinespace
6 & 0.12 & 0.01 & 0.10\\
7 & 0.14 & 0.13 & 0.00\\
8 & 0.28 & 0.28 & 0.03\\
9 & 0.05 & 0.04 & 0.02\\
10 & 0.00 & 0.01 & 0.04\\
\addlinespace
11 & 0.11 & 0.11 & 0.05\\
12 & 0.02 & 0.03 & 0.06\\
13 & 0.09 & 0.01 & 0.09\\
14 & 0.04 & 0.03 & 0.18\\
15 & 0.08 & 0.08 & 0.06\\
\addlinespace
16 & 0.11 & 0.06 & 0.33\\
17 & 0.16 & 0.10 & 0.25\\
18 & 0.17 & 0.02 & 0.34\\
19 & 0.12 & 0.01 & 0.16\\
20 & 0.28 & 0.01 & 0.17\\
\bottomrule
\end{tabular}
\label{tab:asmd_values}
\end{table}

\subsection{Results}

\subsubsection{Anchor-based Method}
\begin{figure}[ht]
    \centering
    \begin{subfigure}[t]{0.49\textwidth}
        \centering
        \includegraphics[width=\textwidth]{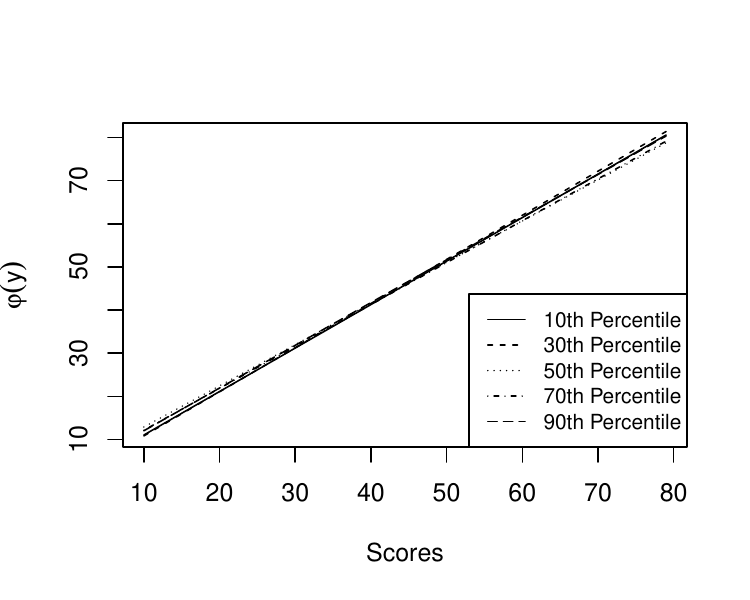}
        \caption{The estimated equated scores for five different anchor scores.}
        \label{fig:equated_scores}
    \end{subfigure}
    \hfill
    \begin{subfigure}[t]{0.49\textwidth}
        \centering
        \includegraphics[width=\textwidth]{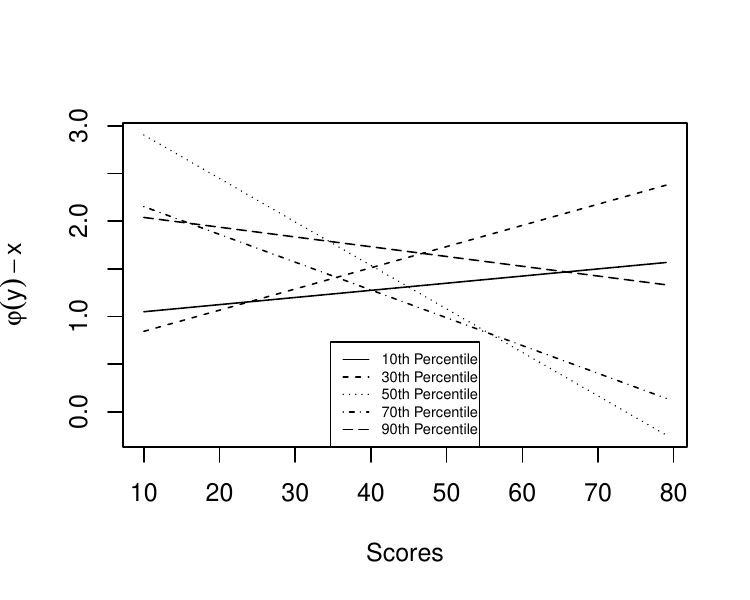}
        \caption{The estimated equated scores, with the unadjusted raw scores subtracted, for five different anchor scores.}
        \label{fig:equated_scores_minus_raw_scores}
    \end{subfigure}
    \caption{The estimated equating functions, conditioning on different values of the anchor score. The 10th percentile corresponds to an anchor score of 5, the 30th percentile to 8, the 50th percentile to 11, the 70th percentile to 14, and the 90th percentile to 18. Panel (a) shows the equated scores, and Panel (b) shows the difference between equated scores and raw scores.}
    \label{fig:side_by_side_equated_scores_anchor}
\end{figure}

Figure \ref{fig:side_by_side_equated_scores_anchor} illustrates two plots of the equated scores using the anchor-based method, based on five selected percentiles of the anchor score. The first plot (a) shows the estimated equating functions, and the second plot visualizes the same equating functions but with the raw scores subtracted from each equated value to highlight the differences between the function values.

The equating functions are relatively close to each other, indicating that the differences between the selected percentiles of the anchor do not result in large differences across most of the score range. This suggests that the examinees from different anchor percentiles follow a relatively consistent pattern of score distribution. In the second plot (b), the equating functions minus the raw scores are displayed. The plot reveals that for lower and higher scores, the equated values exhibit some variation depending on the anchor score percentile. The equated scores conditioning on the 50th and 70th anchor percentile are slightly more similar to each other compared to the other lines, as are the equated scores conditioning on the 10th and 30th anchor percentile. For the mid-range score values, the equated scores for all considered anchor scores are fairly similar.

\subsubsection{Propensity Score Stratification Method}

\begin{figure}[ht!]
    \centering
    \begin{subfigure}[t]{0.49\textwidth}
        \centering
        \includegraphics[width=\textwidth]{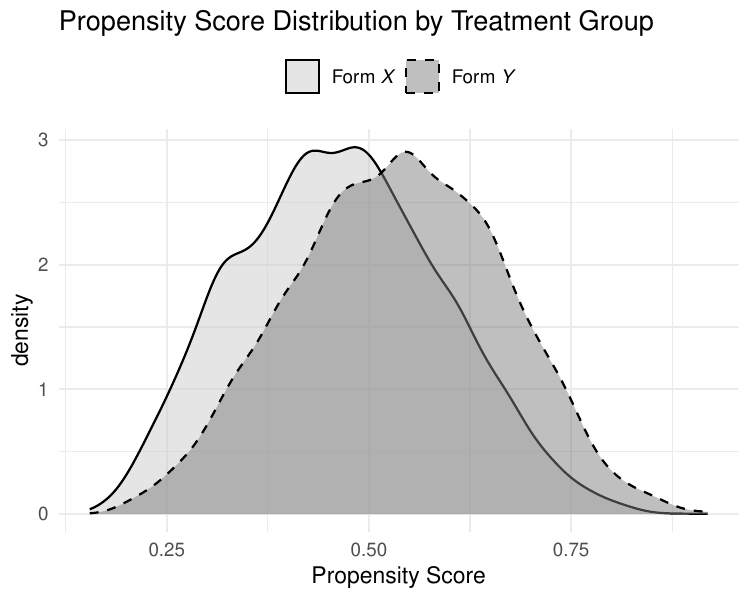}
        \caption{The estimated density function of the estimated propensity scores for each test group. }
        \label{fig:density_ps}
    \end{subfigure}
    \hfill
    \begin{subfigure}[t]{0.49\textwidth}
        \centering
        \includegraphics[width=\textwidth]{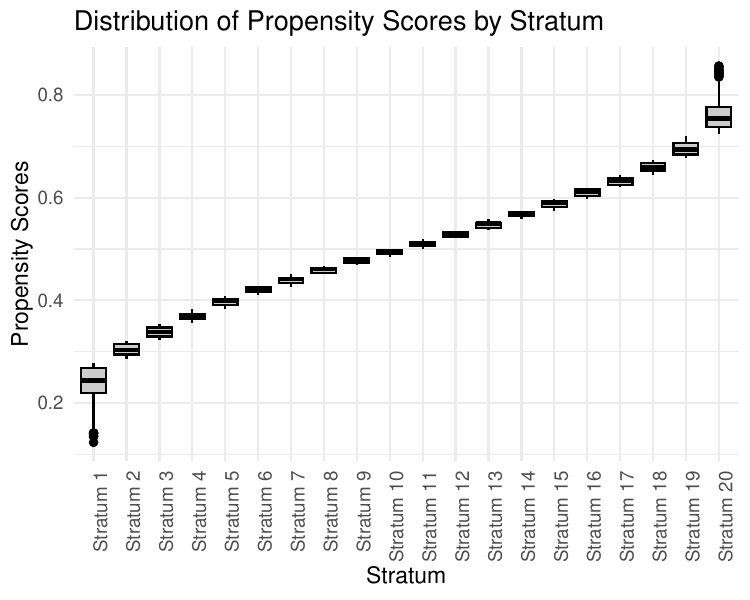}
        \caption{The estimated and stratified propensity scores across 20 strata.}
        \label{fig:ps_boxplot}
    \end{subfigure}
    \caption{The estimated propensity scores.}
    \label{fig:side_by_side_propensity_score}
\end{figure}

In Figure \ref{fig:side_by_side_propensity_score}, the distribution of propensity scores within the treatment groups and across different strata are illustrated. The first plot (a) shows the distribution of propensity scores for the two test groups. There is a clear overlap between the groups in terms of the propensity score distributions. The second plot (b) illustrates boxplots of propensity scores across 20 strata.

\begin{figure}[ht!]
    \centering
    \begin{subfigure}[t]{0.49\textwidth}
        \centering
        \includegraphics[width=\textwidth]{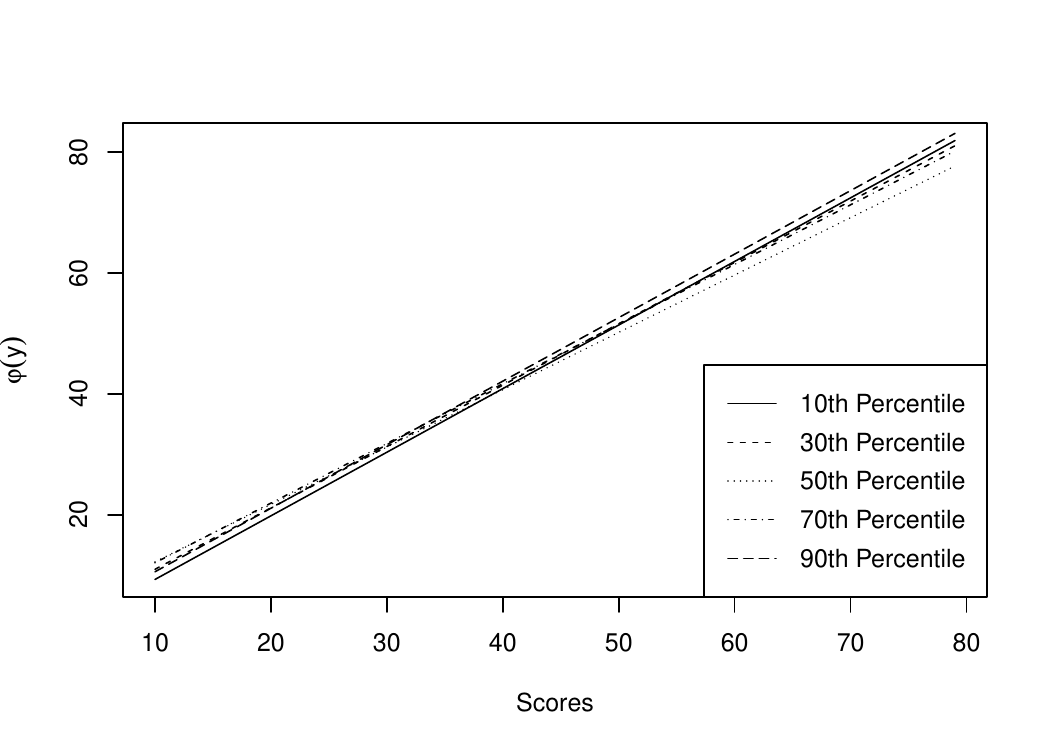}
        \caption{The estimated equated scores for five different estimated and stratified propensity scores.}
        \label{fig:equated_scores_ps}
    \end{subfigure}
    \hfill
    \begin{subfigure}[t]{0.49\textwidth}
        \centering
        \includegraphics[width=\textwidth]{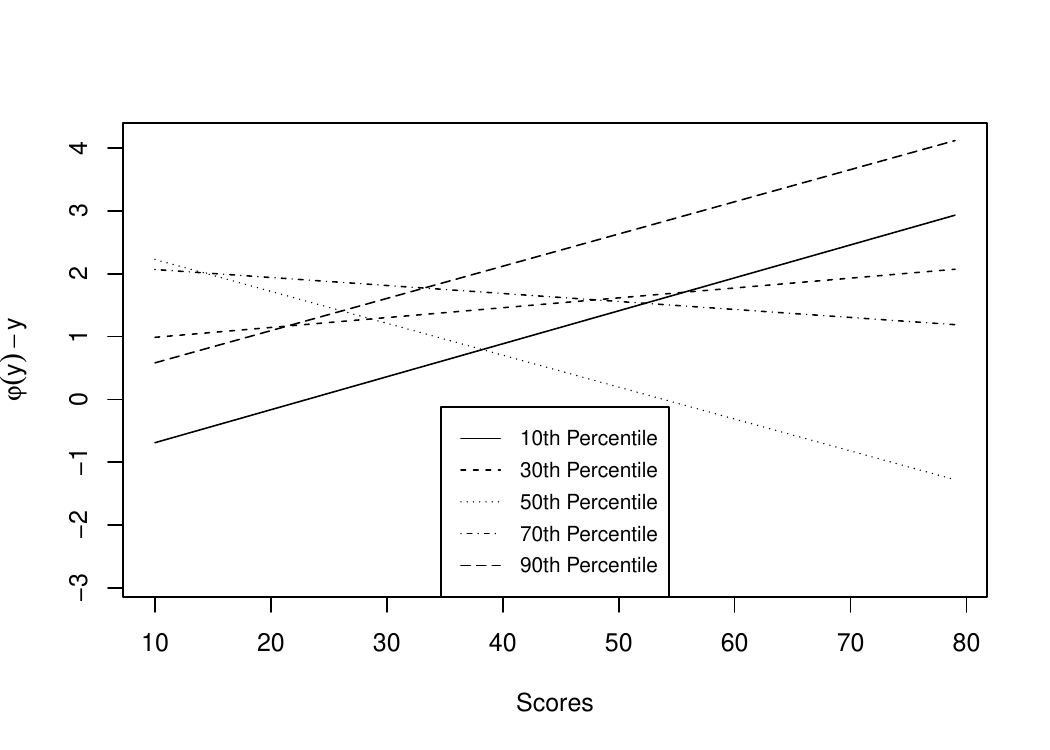}
        \caption{The estimated equated scores, with the unadjusted raw scores subtracted, for five different estimated and stratified propensity scores.}
        \label{fig:equated_scores_minus_raw_scores_ps}
    \end{subfigure}
    \caption{The estimated equating functions, conditioning on different values of the estimated and stratified propensity score. The 10th percentile corresponds to an anchor score of 2, the 30th percentile to 6, the 50th percentile to 10, the 70th percentile to 14, and the 90th percentile to 18. Panel (a) shows the equated scores, and Panel (b) shows the difference between equated scores and raw scores.}
    \label{fig:side_by_side_ps_based_equating}
\end{figure}

In Figure \ref{fig:side_by_side_ps_based_equating}, the estimated equated scores for the equating method based on propensity scores stratification for five selected percentiles of the stratified propensity scores are illustrated. The left-hand side shows the equated scores, and the right-hand side the equated scores with the raw scores subtracted. Similar with the anchor-based method, the equated scores differ based on what value of the propensity score we condition on. From figure (b), it is evident that the equated scores conditioning on the 30th, 50th, and 90th percentile of stratified propensity score are very similar, and that the equated scores for the 10th and 70th percentile are more similar, although still clearly different.

\subsubsection{IPW Method}

\begin{figure}[ht!]
    \centering
    \includegraphics[width=0.6\textwidth]{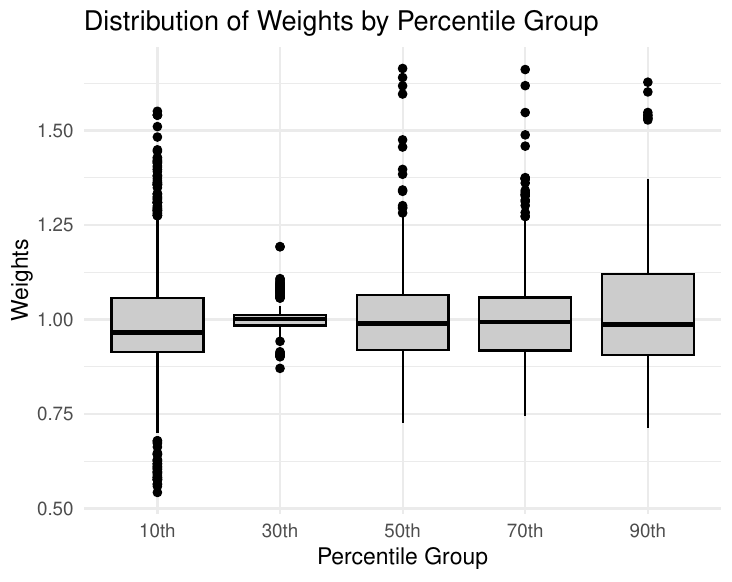}
    \caption{The distribution of the weights used in the IPW-based equating method, across five groups defined by the 10th, 30th, 50th, 70th, and 90th percentile of the estimated and stratified propensity score.}
    \label{fig:weight_distribution}
\end{figure}

In Figure \ref{fig:weight_distribution}, the distribution of the weights used in the IPW-based estimator are displayed for five selected percentiles. The median for all percentile groups are close to 1, indicating that little correction was necessary. Specifically, weights close to one mean that those examinees have estimated propensity scores that align well with their treatment (i.e., test form) assignment. For the 10th percentile group, the weights cover a quite wide range with examinees being both up-weighted and down-weighted. Most examinees have weights in the range approximately 0.75 to 1.25.

In the 30th percentile group, the distribution of weights remains tightly centered around 1, with limited spread. This suggests that for this group, the propensity scores closely match the treatment assignment probabilities, resulting in minimal reweighting. For the 50th, 70th, and 90th percentile groups, they show a broader range of values, with some weights considerably higher than 1, indicating more substantial up-weighting of certain examinees. This increased spread reflects a greater variability in the propensity scores relative to treatment assignment in these groups. The range also suggests that these groups include examinees whose propensity to receive treatment deviates more from what is observed in the data.

\begin{figure}[ht!]
    \centering
    \begin{subfigure}[t]{0.49\textwidth}
        \centering
        \includegraphics[width=\textwidth]{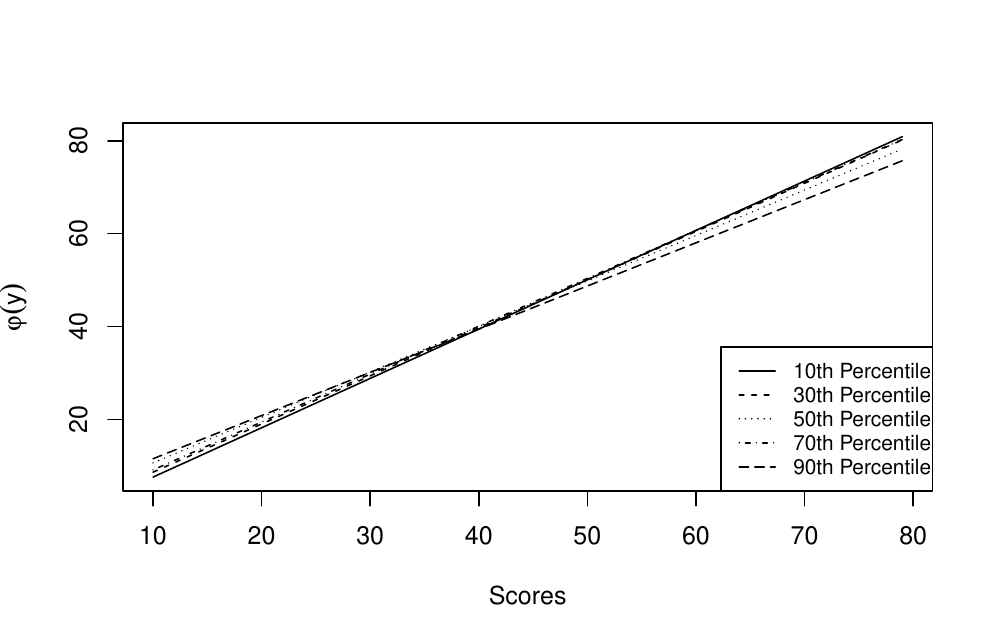}
        \caption{The estimated equated scores for five different estimated and stratified propensity scores.}
        \label{fig:equated_scores_ipw}
    \end{subfigure}
    \hfill
    \begin{subfigure}[t]{0.49\textwidth}
        \centering
        \includegraphics[width=\textwidth]{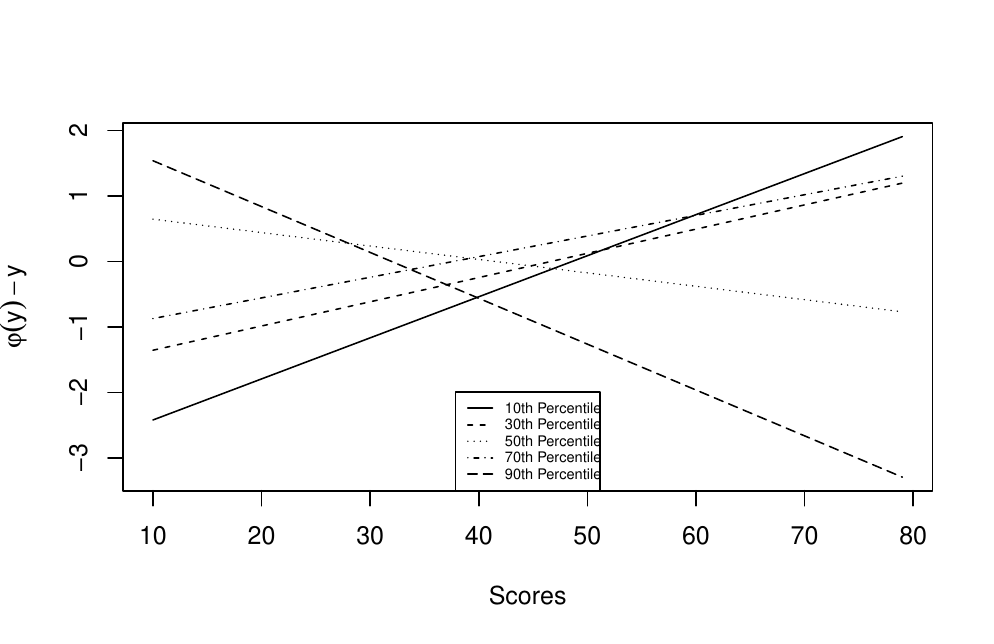}
        \caption{The estimated equated scores, with the unadjusted raw scores subtracted, for five different estimated and stratified propensity scores.}
        \label{fig:equated_scores_minus_raw_scores_ipw}
    \end{subfigure}
    \caption{The estimated IPW-based equating functions, conditioning on different values of the estimated and stratified propensity score. The 10th percentile corresponds to an anchor score of 2, the 30th percentile to 6, the 50th percentile to 10, the 70th percentile to 14, and the 90th percentile to 18. Panel (a) shows the equated scores, and Panel (b) shows the difference between equated scores and raw scores.}
    \label{fig:side_by_side}
\end{figure}

In Figure \ref{fig:equated_scores_minus_raw_scores_ipw}, the estimated equated scores are illustrated for the IPW-based method. As for the two other methods, the equating transformations are very similar in the mid-range of the score scale, but differ clearly in the lower and higher ends. The equated scores, when conditioning on the 30th and 70th percentile of the propensity score, are close to each other, and the 10th percentile curve has a similar slope, whereas the 50th and 90th percentile lines are both negative.

\subsubsection{Comparison Between the Methods}

\begin{figure} [h!]
    \centering 
    \includegraphics[width=1\linewidth]{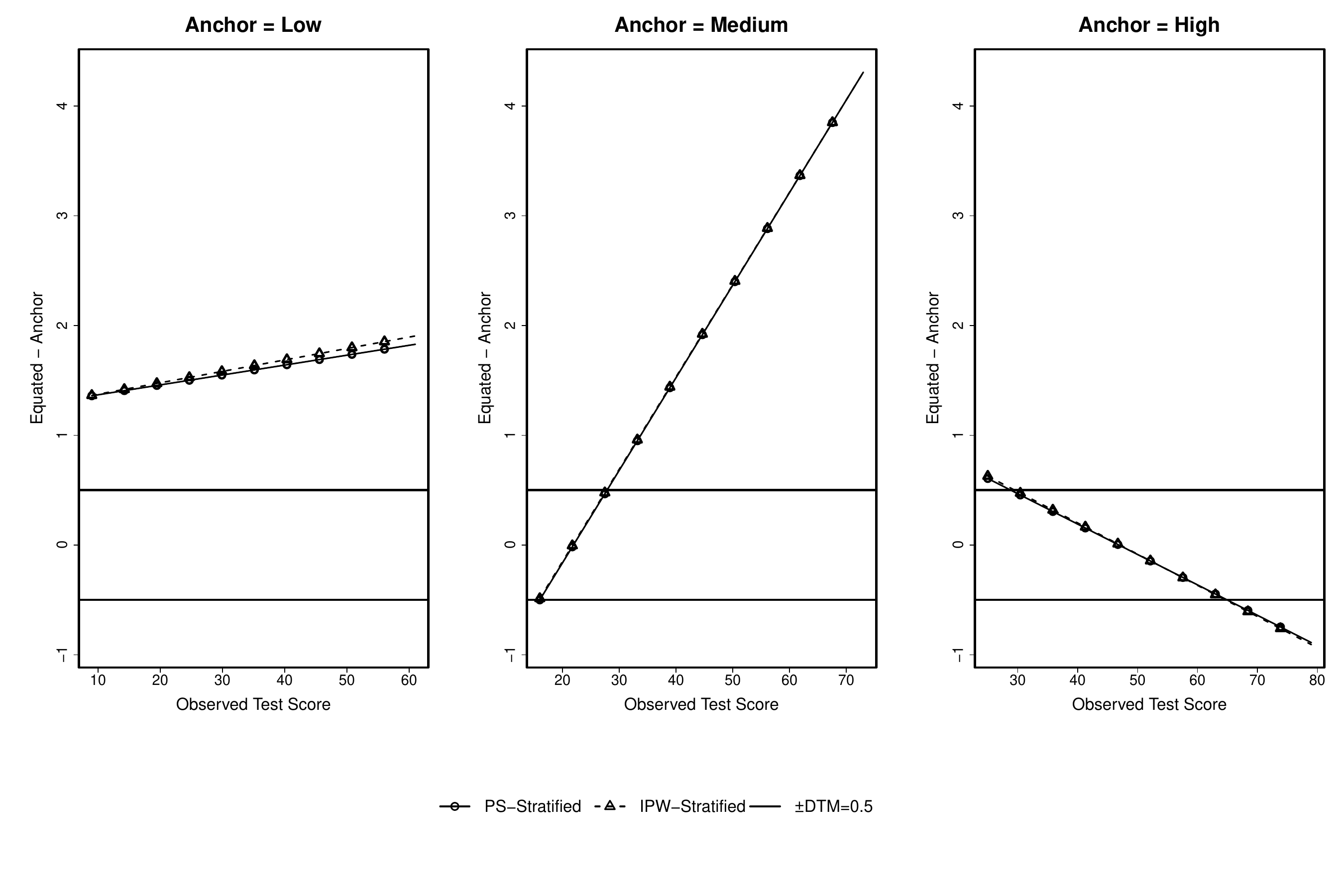}
    \caption{The difference in equated scores between the propensity score-based methods and the anchor-based method, computed within anchor-defined strata (Low, Medium, High). The horizontal solid lines represent the DTM threshold.}
    \label{fig:Equating_Differences}
\end{figure}

In Figure \ref{fig:Equating_Differences}, we compare the anchor-based and propensity score-based equating functions within anchor-defined strata. Specifically, the sample was partitioned into three groups based on anchor scores (Low = 0–33\%, Medium = 33–67\%, High = 67–100\%), and separate anchor-based equating functions were estimated within each stratum. We then evaluated the corresponding propensity score–based equating functions in the stratum whose median estimated propensity score fell within the same anchor tertile, so that both approaches are applied under parallel conditions. 
The vertical axis represents the equated score minus the anchor‐based score; the horizontal axis is the observed test score. In each panel, the line with circle markers depicts the stratified propensity‐score equating function evaluated in the stratum whose median estimated propensity score falls within the same anchor tertile; the line with triangle markers depicts the IPW equating function in that stratum. Horizontal solid lines indicate the Difference That Matters (DTM) threshold for number‐correct scoring \citep{dorans1994equating}.

Both propensity‐score–based estimators yield nearly identical results, with deviations remaining small across the full score range in all three tertiles. In the High anchor tertile, equated scores fall within the DTM bounds for the vast majority of observed scores. By contrast, in the Low and Medium anchor tertiles, differences exceed the DTM threshold across much of the score scale, indicating that the difference between the propensity score-based and anchor score-based equating methods are the greatest when anchor difficulty is at the lower or intermediate levels. This pattern is due to the fact that examinees in the High anchor tertile are more homogeneous in both ability and covariates, so the anchor‐based and propensity score–based methods operate under nearly equivalent conditions. At lower anchor levels, greater heterogeneity remains, leading to larger discrepancies between the two approaches.

\section{Simulation Study}

\subsection{Design}
To evaluate the performance of the proposed equating methods, we conducted a comprehensive simulation study. We generated item response data using a two-parameter logistic (2PL) model, where the probability of a correct response for examinee $i$ on item $k$ is given by 
$$
P(X_{il} = 1 | \theta_i, a_l, b_l) = [1 + \exp(-a_l(\theta_i - b_l))]^{-1},
$$ 
with $\theta_i$ representing the latent ability of examinee $i$, $a_l$ the discrimination parameter, and $b_l$ the difficulty parameter for item $l$. We simulated ability distributions according to $\theta_P \sim N(0, 1)$ and $\theta_Q \sim N(0.5, 1)$. Item parameters were generated to reflect realistic test characteristics: discrimination parameters $a_l \sim U(0.5, 2)$ and difficulty parameters $b_l \sim N(0, 1)$. These parameter ranges are consistent with those used in prior simulation studies (see e.g., \cite{andersson2017item} and reflect typical item behavior in operational tests. The total number of items was fixed at 60, with 20 common items shared between the two forms ($\mathcal{X}$ and $\mathcal{Y}$). This proportion of anchor items approximately aligns with our empirical analysis (33.33\% and 31.25\%, respectively). The sample size ($N$) was varied according to $N=\{1000, 5000\}$. Although test forms $\mathcal{X}$ and $\mathcal{Y}$ were chosen to be smaller than the empirical illustration this conservative choice allows us to safely generalize the results to longer tests, as concluded by \citet{wiberg;vanderlinden;2011}.

We included three covariates, each intended to represent observable characteristics commonly available in empirical testing contexts, such as age group, gender, or scores in other domains. These types of variables are typically reported in ordered categorical form (e.g., age bands, educational attainment levels, grouped scores), and our simulation design aims to reflect this structure. To generate these covariates in a way that captures their potential relationship with latent ability, each was constructed by summing a small number of binary indicators. For example, a covariate with four categories was generated by summing three binary variables, each simulated using a 2PL model conditional on $\theta_i$, with indicator-specific discrimination parameters $a_c \sim \text{Uniform}(0.5, 1.5)$ and difficulty parameters $b_c \sim \mathcal{N}(0, 1)$. The sum yields an integer-valued, ordinal covariate with values ranging from 0 to the number of binary components. This approach allows us to control the strength of association between each covariate and latent ability through the choice of the underlying item parameters. For comparison, we also examined a weaker correlation scenario by drawing $a_c \sim \text{Uniform}(0.1, 0.5)$.

We note that this strategy differs from using a polytomous item response model (e.g., the graded response model). Instead of simulating a single categorical response, we model a collection of binary variables whose outcomes depend on ability, then aggregate them to obtain the covariate. This method yields observed covariates with an ordinal structure, as seen in practice, introduces a tunable correlation between the covariate and latent ability, and allows the covariates to function as observable proxies for the unmeasured confounding induced by $\theta_i$. 


True propensity scores were generated using a logistic model incorporating both the standardized anchor test score ($A$) and the three standardized covariates ($C_1$, $C_2$, $C_3$):
$$
p = [1 + \exp(-(\beta_0 + \beta_A A + \beta_1 C_1 + \beta_2 C_2 + \beta_3 C_3))]^{-1},
$$
where $\beta_0 = 0.0$, $\beta_A = -0.35$, $\beta_1 = 0.1$, $\beta_2 = -0.1$, and $\beta_3 = 0.1$. These coefficients were deliberately chosen to achieve an approximately balanced assignment of individuals to the two test forms, with $E(p) \approx 0.5$. They also introduce moderate dependencies between the covariates, anchor scores, and form assignment, reflecting plausible selection patterns in applied testing contexts. The number of strata was determined through an evaluation of covariate balance, measured by the absolute standardized mean difference (ASMD) within each stratum. After examining various stratification schemes, we fixed the number of strata at 8 for all scenarios.

This design allows us to evaluate the equating methods under various realistic testing scenarios. We emphasize that the true latent abilities $\theta$ were used solely to define population-level equating functions and to evaluate method performance (e.g., conditional bias and RMSE), but not in the implementation of the equating methods themselves. All methods relied only on observed test scores, anchor scores, and covariates as inputs. Therefore, no scale linking was required, and no IRT-based ability estimates were used in practice.

Three equating methods were compared: the traditional anchor-based equating method, the propensity score stratification equating method, and the IPW equating method. We conducted 500 replications for each scenario considered. The simulation study was carried out in R \citep{R;2024}, and the code can be obtained upon request from the corresponding author.

\subsection{Evaluation Measures}

For each simulation replication, we sample examinees from both populations, extracting their total scores, anchor scores, covariates scores, and latent abilities. Based on the sampled data, we estimate the equating transformations for the three methods: anchor-based equating, propensity score stratification, and IPW. To evaluate bias, we partition the sample of examinees from test form $\mathcal{Y}$ into bins based on the observed values of \(\theta_i\), denoted as \(\{\theta_j\}_{j=1}^{\text{nbins}}\), where each bin represents a subinterval \([\theta_j, \theta_{j+1}]\) of the range of \(\theta\). For each examinees in test form $\mathcal{Y}$ whose ability falls within the interval \([\theta_j, \theta_{j+1}]\), the corresponding equated score \(\hat{\varphi}_M(Y_i)\) is computed using the equating transformation for method \(M\) (anchor-based, propensity score stratification, or IPW). The equating transformations \(\hat{\varphi}_M(Y_i)\) transform scores from test form $\mathcal{Y}$ to the scale of test form $\mathcal{X}$ based on the specific proxy used by each method: the anchor score for anchor-based equating, the estimated propensity score strata for propensity score stratification, and the stabilized IPW weights for the IPW method.

The true equated score for an examinee in bin \([\theta_j, \theta_{j+1}]\) is derived from the population-level distributions of \(X\) and \(Y\). Specifically, let \(\mu_{X\mid\theta_j}\) and \(\mu_{Y\mid\theta_j}\) denote the conditional means of the total scores for forms $\mathcal{X}$ and $\mathcal{Y}$, respectively, given that the examinee's latent ability lies in the bin \([\theta_j, \theta_{j+1}]\). Similarly, let \(\sigma_{X\mid\theta_j}\) and \(\sigma_{Y\mid\theta_j}\) denote the conditional standard deviations of the scores on test forms $\mathcal{X}$ and $\mathcal{Y}$ given \(\theta_j\). The true equating transformation is then:
\[
\varphi^*(Y_i) = \mu_{X\mid\theta_j} + \frac{\sigma_{X\mid\theta_j}}{\sigma_{Y\mid\theta_j}}(Y_i - \mu_{Y\mid\theta_j}),
\]
which transforms the observed score \(Y_i\) from test form $\mathcal{Y}$ to the scale of test form $\mathcal{X}$, accounting for the differences in the conditional distributions of scores given \(\theta\). To calculate the bias for each method, we compute the expected absolute difference between the equated score \(\hat{\varphi}_M(Y_i)\) produced by method \(M\) and the true equated score \(\varphi^*(Y_i)\), given \(\theta_j\). The bias for method \(M\) in bin \([\theta_j, \theta_{j+1}]\) is defined as:
\[
\text{Bias}_M(\theta_j) = E\left[\left|\varphi^*(Y_i) - \hat{\varphi}_M(Y_i)\right| \,\middle|\, \theta_j\right],
\]
where the expectation is taken over the distribution of the observed score \(Y_i\) and the proxy (anchor score, propensity score stratum, or IPW weights), conditional on \(\theta_j\). This bias quantifies the average deviation of the equated scores produced by method \(M\) from the true equated scores across different levels of latent ability \(\theta\).

To estimate the bias in practice, we approximate \(E\left[\left|\varphi^*(Y_i) - \hat{\varphi}_M(Y_i)\right| \,\middle|\, \theta_j\right]\) using the average absolute differences over multiple Monte Carlo replications. Specifically, for each replication \(m\), we compute the equated scores \(\hat{\varphi}_M^{(m)}(Y_i)\) for all examinees within each \(\theta\) bin using the equating transformation for each method. The final bias for each method is obtained by averaging the absolute differences over all replications and examinees within each \(\theta\) bin:
\[
\widehat{\text{Bias}}_M(\theta_j) = \frac{1}{\text{MC}_{\text{reps}}} \sum_{m=1}^{\text{MC}_{\text{reps}}} \frac{1}{n_{\theta_j}} \sum_{i \in \theta_j} \left|\varphi^*(Y_i) - \hat{\varphi}_M^{(m)}(Y_i)\right|,
\]
where \(n_{\theta_j}\) is the number of examinees in bin \([\theta_j, \theta_{j+1}]\).

In addition to bias, we evaluate the Root Mean Squared Error (RMSE) of the equating methods. The RMSE for method \(M\) in bin \([\theta_j, \theta_{j+1}]\) is defined as:
\[
\text{RMSE}_M(\theta_j) = \sqrt{ E\left[ \left( \varphi^*(Y_i) - \hat{\varphi}_M(Y_i) \right)^2 \,\middle|\, \theta_j \right] },
\]
where the expectation is again taken over the distribution of \(Y_i\) and the proxy, conditional on \(\theta_j\). 

To estimate the RMSE in practice, we approximate the expectation using the average squared differences over multiple Monte Carlo replications. The RMSE for each method is obtained by computing the square root of the mean squared differences between the true equated scores and the estimated equated scores over all replications and examinees within each \(\theta\) bin:
\[
\widehat{\text{RMSE}}_M(\theta_j) = \sqrt{ \frac{1}{\text{MC}_{\text{reps}}} \sum_{m=1}^{\text{MC}_{\text{reps}}} \frac{1}{n_{\theta_j}} \sum_{i \in \theta_j} \left( \varphi^*(Y_i) - \hat{\varphi}_M^{(m)}(Y_i) \right)^2 }.
\]

\subsection{Results}

\begin{sidewaysfigure}
    \centering
    \includegraphics[width=1\textwidth]{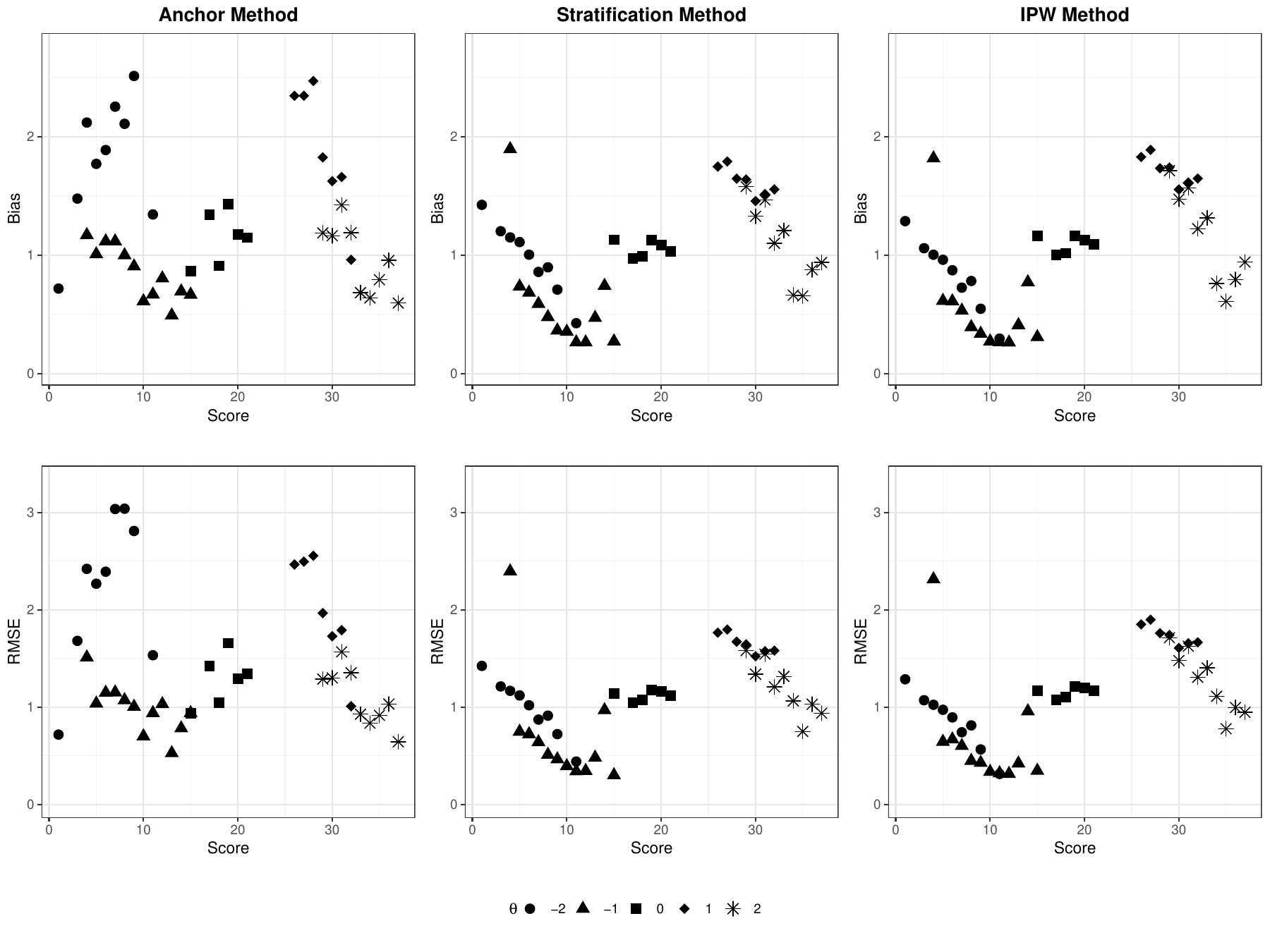}
    \caption{The performance in terms of average bias and RMSE of anchor-based, propensity score-based, and IPW-based equating methods across sum scores for selected theta values for \(N = 1000\) and \(J = 40\) when the correlations between the covariates and the test scores are of weak strength.}
    \label{fig:side_by_side_N1000_weak_corr}
\end{sidewaysfigure}

Figure \ref{fig:side_by_side_N1000_weak_corr} shows the bias and RMSE under the weak correlation scenario, for selected $\theta$ values across the score range, comparing the performance of the three considered equating methods. What is immediately noticeable is that the bias and RMSE are not displayed for all score values. The reason is that all results for score values with a probability lower than $10^{-4}$ have been omitted as their equatings are not considered stable enough. In practice, this choice does not have any consequences since only very few examinees would obtain one of the omitted values. These findings align well with what was found in \citet{wiberg;vanderlinden;2011}, who pointed out that these scores should never be reported without a warning about their lack of accuracy. As noted by these authors, such scores fall in regions where the equating transformation becomes highly unreliable, making unconditional reporting problematic. In practice, this choice affects very few test takers since only a very small number would obtain one of these extreme values, but we argue that the potential for substantial equating error in these regions justifies their exclusion from routine reporting.

In Figure \ref{fig:side_by_side_N1000_weak_corr}, the three methods yield very similar results across most of the score range, especially the propensity score stratification and IPW methods, which are nearly indistinguishable in both bias and RMSE. The average absolute difference between these two methods is only 0.08 for both metrics, with no substantial divergence at any particular score level. The anchor method also performs similarly in many regions. Differences emerge for certain scores, especially at the lower end of the score range where the anchor method exhibits higher bias and RMSE. Unlike the propensity score-based methods, which directly account for covariate information, the anchor method relies solely on the assumption that examinees with the same anchor scores have similar abilities across forms. When the correlation between covariates and ability weakens, the conditional distribution of abilities given anchor scores may differ more substantially across the two populations, particularly at the extremes of the score distribution. This explains why the anchor method's performance is slightly worse at the lower end of the score range, where sample sizes are typically smaller and estimation is inherently more challenging.
Interestingly, despite not directly incorporating covariates in its equating procedure, the anchor method's performance is indirectly affected by covariate correlation changes through the form assignment mechanism. When covariates have weaker correlations with ability, the propensity model for form assignment becomes more heavily influenced by the anchor score alone, potentially creating more systematic differences in ability distributions between forms that are inadequately captured by the anchor items. This highlights an important characteristic of non-equivalent group equating designs: the anchor method's effectiveness depends not only on the anchor items themselves but also on the underlying mechanisms determining sample selection into different forms.


\begin{sidewaysfigure}
    \centering
    \includegraphics[width=1\textwidth]{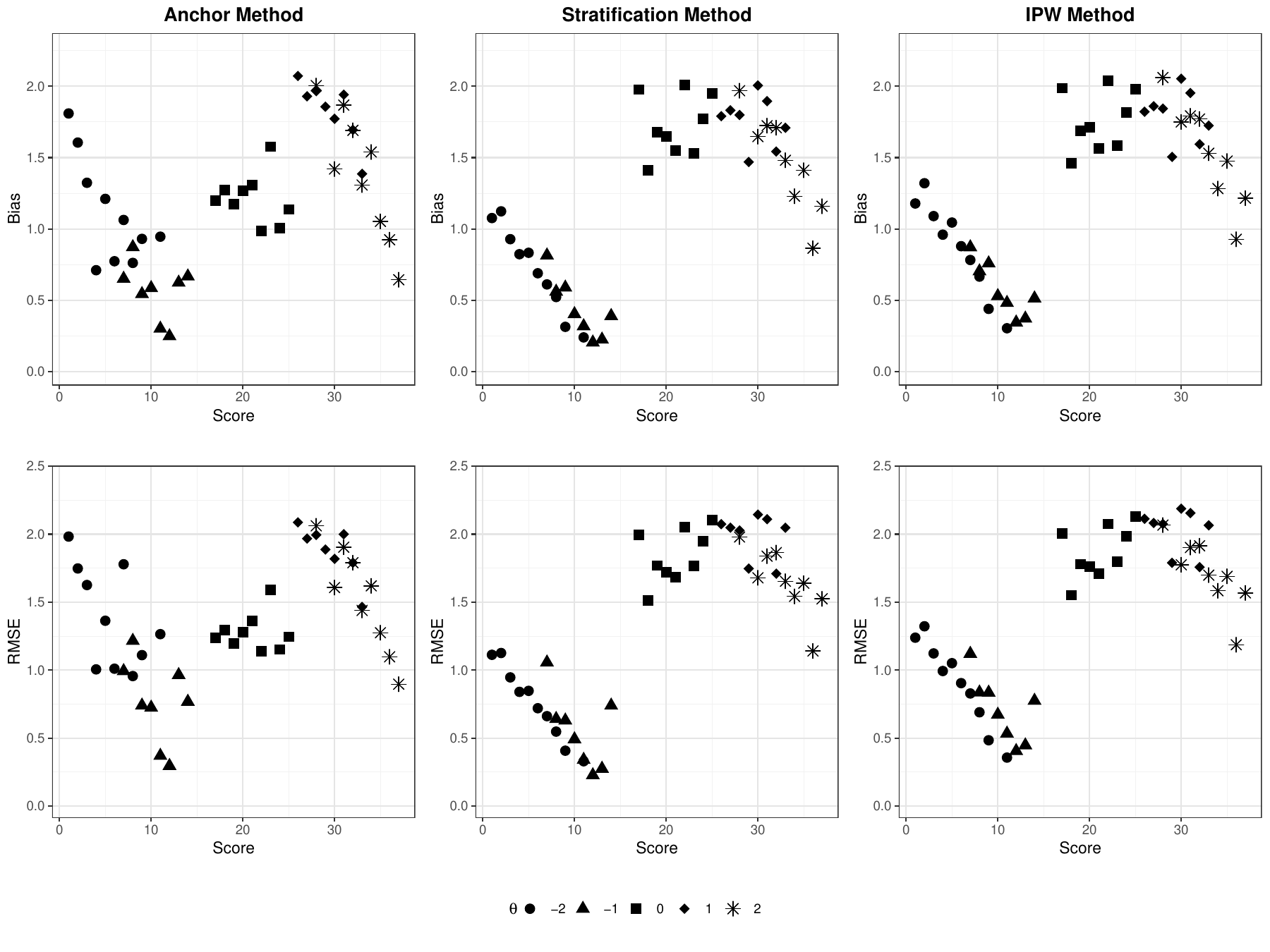}
    \caption{The performance in terms of average bias and RMSE of anchor-based, propensity score-based, and IPW-based equating methods across sum scores for selected theta values for \(N = 1000\) and \(J = 40\) when the correlations between the covariates and the test scores are of medium strength.}
    \label{fig:side_by_side_N1000_medium_corr}
\end{sidewaysfigure}

In Figure \ref{fig:side_by_side_N1000_medium_corr}, panel (a) shows the bias for the anchor, propensity score stratification, and IPW equating methods under the medium correlation scenario. The results are similar to the weak correlation setting, however, both bias and RMSE are slightly smaller which is to be expected.

As mostly similar results were obtained when $N=5000$, these figures are displayed in the Appendix. 


\section{Discussion}

The objective of this study was to explore the use of covariates in local test score equating, particularly when no anchor test is available. We introduced two propensity score-based methods - propensity score stratification and IPW - as alternatives to traditional anchor-based equating methods. Through both empirical and simulation studies, we aimed to assess the effectiveness of these methods in producing fair and comparable test scores across different test forms administered to non-equivalent groups.

From the empirical analysis using data from the SweSAT, we observed that both propensity score-based methods produced equated scores that varied depending on the percentile of the propensity score conditioned upon. This result is in line with the results of previous local equating studies when other proxies have been used to capture differences between group differences \citep[e.g.][]{van2010local,wiberg;vanderlinden;2011,wiberg2014local}. This variation was particularly noticeable at the lower and higher ends of the score scale, while the equated scores were relatively similar in the mid-range. The anchor-based method exhibited a similar pattern, suggesting that conditioning on either an anchor score or estimated propensity scores can capture variations in examinee abilities across different score levels. The propensity score stratification method showed equated scores that were closely aligned for the 30th, 50th, and 90th percentiles, indicating consistency in the mid to higher ability ranges. The IPW method also demonstrated stability in the mid-range but showed differences at the extremes of the score distribution.

In the simulation study, we evaluated the performance of the proposed methods under various conditions, varying factors such as the strength of the correlation between covariates and the latent ability, and the sample size. Note that, to examine different levels of correlation has neither been examined in local equating studies \citep[e.g.][]{van2010local,wiberg;vanderlinden;2011}, nor when equating with propensity scores \citep{wallin;wiberg;2019}. The results indicated that when the correlation between covariates and the test scores was weak, the IPW method consistently exhibited smaller bias compared to the anchor-based and propensity score stratification methods across most of the score range. The bias and RMSE for the IPW method were relatively stable, suggesting robustness in scenarios where covariates are not strongly predictive of the latent ability. When the correlation between covariates and test scores was medium, the performance of the anchor-based method improved, showing decreased bias at higher levels of latent ability. The propensity score stratification method exhibited increasing bias and RMSE across the score scale for a given ability level, but performed better than the IPW method. Surprisingly, the IPW method's performance was slightly worse in the medium correlation scenario compared to the weak correlation scenario, although it still maintained relatively low bias and RMSE.

The findings suggest that both the propensity score stratification and IPW methods may offer viable alternatives to anchor-based equating in situations where an anchor test is unavailable but covariates are available. The stability of the propensity score-based methods across different levels of correlation indicates their potential for adjusting for group differences when covariates do not perfectly capture the variations in ability between groups.

The results align with previous research that has explored the use of covariates and propensity scores in equating. \citet{wiberg;branberg;2015} demonstrated that covariate-adjusted equating methods could effectively account for group differences in the NEC design. Our study extends this work by applying propensity score methods within the local equating framework, addressing the challenge of fulfilling Lord's equity requirement in the absence of an anchor test. Additionally, our findings corroborate those of \citet{wallin;wiberg;2019}, who proposed the use of propensity scores in the NEC design and highlighted the potential of these methods to reduce equating bias.

Although local equating procedures may produce different conversion functions for different subgroups the idea of stratified reporting is not new in testing programs as several of them already report scores conditional on subgroup membership (e.g., by grade level, test version, or test-taking language). Also, our proposed local equating methods can be implemented to obtain conversion functions for a manageable number of strata, rather than a separate function for every individual. The conversion tables can be pre-computed and embedded in scoring software, making operational implementation feasible. Finally, as local equating can reduce bias and improve score interpretation in the presence of group differences this can outweigh the added complexity of using local equating.

Despite the promising results in this study, several limitations should be acknowledged. First, the success of propensity score methods relies heavily on the quality and relevance of the covariates used. In practice, covariates must be related to the latent ability to effectively adjust for group differences. If important covariates are omitted or the relationship between covariates and ability is too weak, the equating transformation may be biased. Additionally, the assumption of unconfoundedness - that all relevant covariates have been included and correctly measured - is a strong one and may not hold in all testing scenarios. Thus it is important to perform robustness studies, similar to \citet{wallin;wiberg;2023}.

It is also important to note that relying solely on covariates in the equating procedure carries certain risks. While covariate-based methods can provide a useful alternative when anchor tests are unavailable, they should be considered as a supplementary approach rather than a replacement for anchor-based methods. Covariates may not capture all aspects of the latent ability, and their effectiveness depends on how well they correlate with the constructs being measured. Therefore, testing programs should aim to include anchor items in their tests, especially when test groups tend to be nonequivalent. Incorporating anchor items provides a direct measure to link test forms and can enhance the accuracy and fairness of the equating process.

Our results and other studies considering propensity scores in equating indicate that under certain conditions, covariate-based methods can perform almost as well as, and sometimes even better than, anchor-based methods. This suggests that in situations where anchor items are not available, covariate-based methods serve as a "better than nothing" alternative. Moreover, for testing programs that are beginning to incorporate anchor items, there is potential value in combining information from both covariates and anchor items when performing equating. Such an approach could leverage the strengths of both methods, potentially improving the accuracy of the equating transformation. However, methodologies for integrating both covariates and anchor items in the equating process are yet to be fully developed, and we leave this as a direction for future research.

Future research could also explore optimal stratification strategies or the use of alternative methods, such as propensity score matching, to enhance the balance between groups. Additionally, investigating the impact of different types and numbers of covariates on the performance of propensity score-based methods would provide valuable insights. Assessing the generalizability of these methods across various testing contexts, including different test formats and populations, would further contribute to understanding their applicability.

Moreover, while our study focused on linear equating, there is potential for extending the proposed methods to equipercentile equating, as discussed in Section \ref{generalisation_equipercentile}. Equipercentile equating may provide a more flexible approach, particularly for tests with non-linear score distributions or when higher moments need to be matched. 

In conclusion, this study contributes two propensity score-based methods for local equating when no anchor test is available. The findings suggest that both methods hold promise as alternatives to traditional anchor-based methods. While limitations exist, the exploration of covariate-based equating methods expands the toolkit available to assessment professionals, facilitating the development of fair testing practices. Future research should continue to further develop these methods, address their limitations, and explore their applicability across a broader range of testing scenarios, including the integration of both covariates and anchor items in the equating process.

\clearpage
\bibliography{bibliography}

@article{andersson2017item,
  title={Item response theory observed-score kernel equating},
  author={Andersson, Bj{\"o}rn and Wiberg, Marie},
  journal={Psychometrika},
  volume={82},
  number={1},
  pages={48--66},
  year={2017},
  publisher={Springer}
}

@book{kolen2014test,
  title={Test equating, Scaling and Linking: Methods and practices},
  author={Kolen, Michael J and Brennan, Robert L},
  year={2014},
  publisher={New York: Springer}
}

@book{von2004kernel,
  title={The kernel method of test equating},
  author={von Davier, Alina A and Holland, Paul W and Thayer, Dorothy T},
  year={2004},
  publisher={New York: Springer}
}

@article{gonzalez2017applying,
  title={Applying test equating methods using {R}},
  author={Gonz{\'a}lez, Jorge and Wiberg, Marie},
  journal={Cham: Springer},
  year={2017},
  publisher={Springer}
}

@article{wiberg;branberg;2015,
  title={Kernel equating under the non-equivalent groups with covariates design},
  author={Wiberg, Marie and Br{\"a}nberg, Kenny},
  journal={Applied Psychological Measurement},
  volume={39},
  number={5},
  pages={349--361},
  year={2015},
  publisher={SAGE Publications Sage CA: Los Angeles, CA}
}

@article{branberg2011observed,
  title={Observed score linear equating with covariates},
  author={Br{\"a}nberg, Kenny and Wiberg, Marie},
  journal={Journal of Educational Measurement},
  volume={48},
  number={4},
  pages={419--440},
  year={2011},
  publisher={Wiley Online Library}
}

@article{hernan2006estimating,
  title={Estimating causal effects from epidemiological data},
  author={Hern{\'a}n, Miguel A and Robins, James M},
  journal={Journal of Epidemiology \& Community Health},
  volume={60},
  number={7},
  pages={578--586},
  year={2006},
  publisher={BMJ Publishing Group Ltd}
}

@article{thoemmes2011systematic,
  title={A systematic review of propensity score methods in the social sciences},
  author={Thoemmes, Felix J and Kim, Eun Sook},
  journal={Multivariate behavioral research},
  volume={46},
  number={1},
  pages={90--118},
  year={2011},
  publisher={Taylor \& Francis}
}

@article{vikstrom2017dynamic,
  title={Dynamic treatment assignment and evaluation of active labor market policies},
  author={Vikstr{\"o}m, Johan},
  journal={Labour Economics},
  volume={49},
  pages={42--54},
  year={2017},
  publisher={Elsevier}
}

@article{pais2011socioeconomic,
  title={Socioeconomic background and racial earnings inequality: A propensity score analysis},
  author={Pais, Jeremy},
  journal={Social science research},
  volume={40},
  number={1},
  pages={37--49},
  year={2011},
  publisher={Elsevier}
}

@article{austin2008performance,
  title={The performance of different propensity-score methods for estimating relative risks},
  author={Austin, Peter C},
  journal={Journal of clinical epidemiology},
  volume={61},
  number={6},
  pages={537--545},
  year={2008},
  publisher={Elsevier}
}

@article{huber2015causal,
  title={Causal pitfalls in the decomposition of wage gaps},
  author={Huber, Martin},
  journal={Journal of Business \& Economic Statistics},
  volume={33},
  number={2},
  pages={179--191},
  year={2015},
  publisher={Taylor \& Francis}
}

@article{van2010local,
  title={Local observed-score equating with anchor-test designs},
  author={van der Linden, Wim J and Wiberg, Marie},
  journal={Applied Psychological Measurement},
  volume={34},
  number={8},
  pages={620--640},
  year={2010},
  publisher={SAGE Publications Sage CA: Los Angeles, CA}
}

@book{lord1980applications,
  title={Applications of item response theory to practical testing problems},
  author={Lord, Frederic M},
  year={1980},
  publisher={Hillsdale, NJ: Lawrence Erlbaum Associates}
}

@article{kolen1990does,
  title={Does matching in equating work: A discussion},
  author={Kolen, Michael J},
  journal={Applied Measurement in Education},
  volume={3},
  number={1},
  pages={97--104},
  year={1990},
  publisher={Taylor \& Francis}
}

@article{cook1990equating,
  title={Equating achievement tests using samples matched on ability},
  author={Cook, Linda L and Eignor, Daniel R and Schmitt, Alicia P},
  journal={ETS Research Report Series},
  volume={1990},
  number={1},
  pages={i--58},
  year={1990},
  publisher={Wiley Online Library}
}

@article{wright1993using,
  title={Using the selection variable for matching or equating},
  author={Wright, Nancy K and Dorans, Neil J},
  journal={ETS Research Report Series},
  volume={1993},
  number={1},
  pages={i--22},
  year={1993},
  publisher={Wiley Online Library}
}

@article{longford2015equating,
  title={Equating without an anchor for nonequivalent groups of examinees},
  author={Longford, Nicholas T},
  journal={Journal of Educational and Behavioral Statistics},
  volume={40},
  number={3},
  pages={227--253},
  year={2015},
  publisher={SAGE Publications Sage CA: Los Angeles, CA}
}

@article{cochran1968effectiveness,
  title={The effectiveness of adjustment by subclassification in removing bias in observational studies},
  author={Cochran, William G},
  journal={Biometrics},
  pages={295--313},
  year={1968},
  publisher={JSTOR}
}

@book{imbens2015causal,
  title={Causal inference in statistics, social, and biomedical sciences},
  author={Imbens, Guido W and Rubin, Donald B},
  year={2015},
  publisher={Cambridge university press}
}

@article{dorans2008anchor,
  title={Anchor test type and population invariance: An exploration across subpopulations and test administrations},
  author={Dorans, Neil J and Liu, Jinghua and Hammond, Shelby},
  journal={Applied Psychological Measurement},
  volume={32},
  number={1},
  pages={81--97},
  year={2008},
  publisher={Sage Publications Sage CA: Los Angeles, CA}
}

@article{dorans1994equating,
  title={Equating issues engendered by changes to the SAT and PSAT/NMSQT},
  author={Dorans, NJ and Feigenbaum, MD},
  journal={Technical issues related to the introduction of the new SAT and PSAT/NMSQT},
  pages={91--122},
  year={1994}
}

@article{rosenbaum1983central,
  title={The central role of the propensity score in observational studies for causal effects},
  author={Rosenbaum, Paul R and Rubin, Donald B},
  journal={Biometrika},
  volume={70},
  number={1},
  pages={41--55},
  year={1983},
  publisher={Oxford University Press}
}

@article{livingston1990combination,
  title={What combination of sampling and equating methods works best?},
  author={Livingston, Samuel A and Dorans, Neil J and Wright, Nancy K},
  journal={Applied Measurement in Education},
  volume={3},
  number={1},
  pages={73--95},
  year={1990},
  publisher={Taylor \& Francis}
}

@techreport{paek2006propensity,
  author = {Paek, Insu and Liu, Jinghua and Oh, Hyun Jin},
  title = {Investigation of propensity score matching on linear/nonlinear equating method for the P/N/NMSQT},
  institution = {ETS},
  year = {2006},
  number = {SR-2006-55},
  address = {Princeton, NJ},
  type = {Technical Report}
}

@book{sungworn2009investigation,
  title={An investigation of using collateral information to reduce equating biases of the post-stratification equating method},
  author={Sungworn, N},
  year={2009},
  publisher={Ph. D. Thesis, Michigan State University}
}

@article{moses2010use,
  title={The use of two anchors in nonequivalent groups with anchor test (NEAT) equating},
  author={Moses, Tim and Deng, Weiling and Zhang, Yu-Li},
  journal={ETS Research Report Series},
  volume={2010},
  number={2},
  pages={i--33},
  year={2010},
  publisher={Wiley Online Library}
}

@book{powers2010impact,
  title={Impact of matched samples equating methods on equating accuracy and the adequacy of equating assumptions},
  author={Powers, Sonya Jean},
  year={2010},
  publisher={The University of Iowa}
}

@article{hsu2002exploring,
  title={Exploring the feasibility of collateral information test equating},
  author={Hsu, Tse-Chi and Wu, Kuo-liang and Yu, Jya-Yi Wu and Lee, Ming-Yen},
  journal={International Journal of Testing},
  volume={2},
  number={1},
  pages={1--14},
  year={2002},
  publisher={Taylor \& Francis}
}

@article{liou2001estimating,
  title={Estimating comparable scores using surrogate variables},
  author={Liou, Michelle and Cheng, Philip E and Li, Ming-Yen},
  journal={Applied Psychological Measurement},
  volume={25},
  number={2},
  pages={197--207},
  year={2001},
  publisher={Sage Publications Sage CA: Thousand Oaks, CA}
}

@article{lyren2011consequences,
  title={Consequences of violated equating assumptions under the equivalent groups design},
  author={Lyr{\'e}n, Per-Erik and Hambleton, Ronald K},
  journal={International Journal of Testing},
  volume={11},
  number={4},
  pages={308--323},
  year={2011},
  publisher={Taylor \& Francis}
}

@article{rosenbaum1984reducing,
  title={Reducing bias in observational studies using subclassification on the propensity score},
  author={Rosenbaum, Paul R and Rubin, Donald B},
  journal={Journal of the American statistical Association},
  volume={79},
  number={387},
  pages={516--524},
  year={1984},
  publisher={Taylor \& Francis}
}

@incollection{van2011local,
    author={van der Linden, Wim J},
    title = {Local observed-score equating},
    booktitle = {Statistical models for test equating, scaling, and linking},
    publisher={New York: Springer},
    year = {2011}
}

@article{wiberg2014local,
  title={Local observed-score kernel equating},
  author={Wiberg, Marie and van der Linden, Wim J and von Davier, Alina A},
  journal={Journal of Educational Measurement},
  volume={51},
  number={1},
  pages={57--74},
  year={2014},
  publisher={Wiley Online Library}
}

@Article{wallin;wiberg;2019,
  Title  = {{Kernel equating using propensity scores for non-equivalent groups}},
  Author = {Wallin, Gabriel and Wiberg, Marie},
  Journal= {Journal of Educational and Behavioral Statistics},
  Year    = {2019},
  pages   = {390--414},
  volume  = {44},
  number  = {4}
}

@Article{wiberg;vanderlinden;2011,
  Title  = {Local linear observed-score equating},
  Author = {Wiberg, M. and van der Linden, W. J.},
  Journal = {Journal of Educational Measurement},
  Year   = {2011},
  Pages  = {229--254},
  Volume = {48}
}

@book{wiberg;etal;2025,
    author = {Wiberg, M and Gonzalez, J. and von Davier, A.A.},
    title = {Generalized Kernel Equating with applications in {R}},
    publisher = {Boca Raton, FL: Chapman \& Hall} ,
    year = {2025}
}

@Manual{R;2024,
    title           = {R: A Language and Environment for Statistical Computing},
    author          = {{R Core Team}},
    organization    = {R Foundation for Statistical Computing},
    address         = {Vienna, Austria},
    year            = {2024},
    url             = {https://www.R-project.org/},
}

@article{wallin;wiberg;2023,
  title={Model misspecification and robustness of observed-score test equating using propensity scores},
  author={Wallin, Gabriel and Wiberg, Marie},
  journal={Journal of Educational and Behavioral Statistics},
  volume={48},
  number={5},
  pages={603--635},
  year={2023},
  publisher={Sage Publications Sage CA: Los Angeles, CA}
}

@article{horvitz;thompson;1952,
  title={A generalization of sampling without replacement from a finite universe},
  author={Horvitz, Daniel G and Thompson, Donovan J},
  journal={Journal of the American statistical Association},
  volume={47},
  number={260},
  pages={663--685},
  year={1952},
  publisher={Taylor \& Francis}
}
\end{document}